\documentclass[conference,a4paper]{APSIPA2019}
\usepackage{multirow}
\usepackage[dvipdfmx]{graphicx}
\usepackage{amsmath}
\usepackage[psamsfonts]{amssymb}
\usepackage{amsxtra}
\usepackage{threeparttable}

\renewcommand{\baselinestretch}{1}

\begin{document}

\title{Frequency domain variant of Velvet noise\\ and its application to acoustic measurements}

\author{%
\authorblockN{%
Hideki Kawahara\authorrefmark{1},
Ken-Ichi Sakakibara\authorrefmark{2},
Mitsunori Mizumachi\authorrefmark{3},
Hideki Banno\authorrefmark{4},
Masanori Morise\authorrefmark{5} and
Toshio Irino\authorrefmark{1}
}
\authorblockA{%
\authorrefmark{1}
Wakayama University, Wakayama, Japan \\
E-mail: \{kawahara, irino\}@wakayama-u.ac.jp}
\authorblockA{%
\authorrefmark{2}
Health Science University of Hokkaido, Sapporo, Japan\\
E-mail: quesokis@gmail.com}
\authorblockA{%
\authorrefmark{3}
Kyushu Institute of Technology, Kitakyushu, Japan\\
E-mail: mizumach@ecs.kyutech.ac.jp}
\authorblockA{%
\authorrefmark{4}
Meijo Universitty, Nagoya, Japan\\
E-mail: banno@meijo-u.ac.jp}
\authorblockA{%
\authorrefmark{5}
Meiji University, Tokyo, Japan\\
E-mail: mmorise@meiji.ac.jp}
}

\maketitle
\thispagestyle{empty}

\begin{abstract}
We propose a new family of test signals for acoustic measurements such as impulse response, nonlinearity, and the effects of background noise.
The proposed family complements difficulties in existing families, the Swept-Sine (SS),
pseudo-random noise such as the maximum length sequence (MLS).
The proposed family uses the frequency domain variant of the Velvet noise (FVN) as its building block.
An FVN is an impulse response of an all-pass filter and yields the unit impulse when convolved with the time-reversed version of itself.
In this respect, FVN is a member of the time-stretched pulse (TSP) in the broadest sense. 
The high degree of freedom in designing an FVN opens a vast range of applications in acoustic measurement.
We introduce the following applications and their specific procedures, among other possibilities. 
They are as follows. 
a) Spectrum shaping adaptive to background noise. 
b) Simultaneous measurement of impulse responses of multiple acoustic paths. 
d) Simultaneous measurement of linear and nonlinear components of an acoustic path. 
e) Automatic procedure for time axis alignment of the source and the receiver 
when they are using independent clocks in acoustic impulse response measurement.
We implemented a reference measurement tool equipped with all these procedures.
The MATLAB source code and related materials are open-sourced and placed in a GitHub repository.
\end{abstract}

\section{Introduction}
We introduce a new family of test signals for acoustic measurements.
Acoustic measurements such as impulse response have been using pseudo-random signals such as 
the maximum-length sequence (MLS)\cite{schroeder1979integrated,muller2001transfer} and 
swept sine signals (SS)\cite{aoshima1981jasa,farina2000simultaneous,morise2007warped,ochiai2011impulse} 
which have a constant power spectrum while are temporally spread.
For strictly time-invariant linear systems, they provide the same results.
However, in actual measurements, depending on background noise, non-linearity\cite{voishvillo2004graphing,klippel2006tutorial,temme2008new}, 
temporal variability, and airflow (for example, wind) modulation on sound propagation speed,
their results differ\cite{guidorzi2015impulse}.
We introduce a third family of test signals based on the frequency-domain variants of velvet noise (FVN)\cite{kawahara2018sigmus118,kawahara2018IS}
which was inspired by the original velvet noise (OVN)\cite{jarvelainen2007reverberation}.
The high degree of freedom in FVN design 
opens a vast range of applications in acoustic measurement and provide simple solutions to difficulties in MLS and SS-based method;
such as fragility to time axis warping\cite{Mori2017ast}, and 
complexity in the objective assessment of intermodulation distortions caused by multi-component signals\cite{voishvillo2004graphing,temme2008new}.

The velvet noise (OVN) is a sparse discrete signal which consists of fewer than 20\% of non-zero (1 or -1) elements.
The name ``velvet'' represents its perceptual impression.
It sounds smoother than Gaussian white noise\cite{jarvelainen2007reverberation,valimaki2013ieetr}.
We found that the frequency domain variants of velvet noise (FVN, afterward) provide useful candidates for the excitation source signals of
synthetic speech and singing\cite{kawahara2018sigmus118,kawahara2018IS}.
The proposed FVN is also an impulse response of an all-pass filter\cite{oppenheimerBook}.
In other words, FVN is a TSP (Time Stretched Pulse\cite{aoshima1981jasa}) in the broadest sense.
FVN is a unique TSP because it has a high degree of design flexibility,
which opens vast possibilities in acoustic measurement.
Specifically, a) an efficient simultaneous measurement of 
the multiple acoustic paths without interferences, and its application to
b) a flexible simultaneous measurement of the linear and the nonlinear component
are significant contributions of this article.

This article starts from a brief description of the OVN, followed
by an introduction to the FVN.
Then, the following descriptions of application to acoustic measurements use
typical configurations of measurement to introduce the above mentioned simultaneous measurements.
The following numerical example section, we introduce
results of acoustic measurements of several loudspeaker systems and microphones.
Finally, we discuss other possible applications and relations to SS and MLS-based measurements.
We also presented Appendices for introducing technical details about
how to design the unit FVNs and the test signals.

\section{Velvet noise}
The velvet noise was designed for artificial reverberation algorithms.
It is a randomly allocated unit impulse sequence with minimal impulse density vs.
maximal smoothness of the noise-like characteristics.
Because such sequence can sound smoother than the Gaussian noise,
it is named ``velvet noise.''\cite{jarvelainen2007reverberation}

The velvet noise allocates a randomly selected positive or negative unit pulse at
a random location in each temporal segment\cite{jarvelainen2007reverberation,valimaki2013ieetr}.
Let $T_d$ represent the average pulse interval in samples.
The following equation determines the location of the $m$-th pulse $k_\mathrm{ovn}(m)$.
The subscript ``ovn'' stands for ``Original Velvet Noise.''
It uses two sequences of random numbers $r_1(m),$  and $r_2(m)$ generated from a uniform distribution in $(0, 1)$.
\begin{align}
k_\mathrm{ovn}(m) & = ||m Td + r_1(m) (T_d - 1) || ,
\end{align}
where the rounding function $|| \bullet ||$ returns the nearest integer.
The  following equation determines the value of the signal $s_\mathrm{ovn}(n)$ at discrete time $n$.
\begin{align}\label{eq:sigsin}
s_\mathrm{ovn}(n) & = \left\{
\begin{array}{ll}
2 || r_2(m)|| - 1  & n=  k_\mathrm{ovn}(m)  \\
0  &   \mbox{otherwise}   
\end{array}
\right. .
\end{align}

With the pulse density higher than 3,000 pulses per second, OVN sounds like white Gaussian noise
and provides a smoother impression.
Supplemental media consists of OVN examples.
In the following section, we introduce FVN replicating
the descriptions of our article\cite{kawahara2018IS}.

\section{Frequency domain variant of velvet noise}
The discrete Fourier transform of a velvet noise sequence closely approximates a complex Gaussian random sequence.
The discrete Fourier transform of the filtered velvet noise provides a complex Gaussian noise on the frequency axis with the filter shape weighting.
Using the duality of the frequency and the time of Fourier transform, we apply filtered velvet noise to the design phase of the all-pass filter.
The impulse response of this all-pass filter is the element of the proposed FVN.
The element has the temporally localized envelope and random waveform.
The key design issue is the shape of the function to manipulate the phase.


\subsection{Unit of phase manipulation}
We use a set of cosine series functions for manipulating the phase because it is easy to implement well-behaving localization\cite{nattall1981ieee,kawahara2017interspeechGS}.
This section investigates relations between phase manipulation and the impulse response of the corresponding all-pass filter.
Let $w_p(k, B)$ represent a phase modification function on the discrete frequency domain.
The following equation provides the complex-valued impulse response $h(n; k_c, B)$ of the all-pass filter. 
\begin{align}\label{eq:unitphase}
\!\! h(n; k_c, B) & = \! \frac{1}{K}\! \sum_{k = 0}^{K - 1} \!  \exp\!\left(\frac{2 k n \pi j}{KN} + j w_p(k- k_c, B) \right) ,
\end{align}
where $k_c$ represents the discrete center frequency, and
$B$ defines the support of $w_p(k, B)$ in the frequency domain
(i.e. $w_p(k, B) = 0$ for $|k| > B$).
The symbol of the imaginary unit is $j = \sqrt{-1}$, and
$N$ represents the number of DFT bins.

We tested four types of cosine series.
They are Hann, Blackman, Nuttall, and the six-term cosine series used in\cite{kawahara2017interspeechGS}.
The Nuttall's reference\cite{nattall1981ieee} provides a list of coefficients of the first three functions and the design procedure.
The following cosine series defines these functions.
Let define $B_w = B /M$ as nominal bandwidth.
\begin{align}
w_p(k, B) & = \sum_{m = 0}^{M} a(m) \cos\left(\frac{\pi k m}{B} \right) ,
\end{align}
where $M$ represents the highest order of the cosine series.

We also tested the theoretically the best bounded-function, prolate spheroidal wave function\cite{slepian1961prolate}
and its approximation, Kaiser window\cite{kaiser1980use}.
We found that the six-term cosine series provides the best localization behavior.
The six-term series has practically no interference due to sidelobes.
We decided to use this six-term series afterward.
The coefficients of the six-term series are
0.2624710164, 0.4265335164, 0.2250165621, 0.0726831633, 0.0125124215, and 0.0007833203 from $a_0$ to $a_5$.
\footnote{The coefficients are the result of the exhaustive search with ten digits.
Practically, truncating the numbers to six digits does not degrade the results.}
The sidelobes have the highest level of -114~dB and  the decay rate of -54~dB/oct. 

\subsection{Phase manipulation unit allocation using velvet noise}
By adding unit phase manipulation $w_p(k- k_c, B)$ on
a set of center frequencies $k_c$ obeying the design rule of velvet noise yields the
filtered velvet noise in the frequency domain.
The following equation defines the allocation index $k_c = k_\mathrm{fvn}(m)$
where subscript ``fvn'' stands for Frequency-domain variants of Velvet Noise.
\begin{align}
k_\mathrm{fvn}(m) & = ||m F_d + r_1(m) (F_d - 1) || ,
\end{align}
where $F_d$ represents the average frequency segment length.
(Note that the index $k_\mathrm{fvn}(m)$ has a real value in MATLAB implementation, instead of an integer value to 
avoid side effects of quantization.)
Each location spans from 0~Hz to $f_s/2$.
Let $\mathbb{K}$ represent a set of allocation indices $k_\mathrm{fvn}(m)$.
The following equation provides the phase $\varphi_\mathrm{fvn}(k)$ of this frequency variant of velvet noise.
\begin{align}
\!\! \varphi_\mathrm{fvn}(k) & = \!\!\sum_{k_c \in \mathbb{K}} \!\! s_\mathrm{fvn}(k_c) \left(w_p(k\!-\! k_c, B) \!- \!w_p(k\!+\! k_c, B)\right) , \label{eq:vpAlloc} \\
\!\! s_\mathrm{fvn}(m) & =  \left(2 || r_2(m)|| - 1 \right) \varphi_\mathrm{max} 
\end{align}
where $k$ spans discrete frequency of a DFT buffer, which has a circular discrete frequency axis, and
the parameter $\varphi_\mathrm{max}$ defines the magnitude of phase manipulation.
The second term inside of parentheses of Eq.~\ref{eq:vpAlloc} is to make the phase function have the odd symmetry concerning 0~Hz and $f_s/2$.

The inverse discrete Fourier transform of this all-pass filter provides an impulse response.
It is the unit signal $h_\mathrm{fvn}(n)$ of the proposed FVN.
\begin{align}\label{eq:unitfvn}
h_\mathrm{fvn}(n) & = \frac{1}{K}\sum_{k = 0}^{K - 1}  \exp\left(\frac{2 k n \pi j }{KN} + j \varphi_\mathrm{fvn}(k) \right) .
\end{align}

\subsection{FVN design}
\begin{figure}[tbp]
\begin{center}
\includegraphics[width=0.48\hsize]{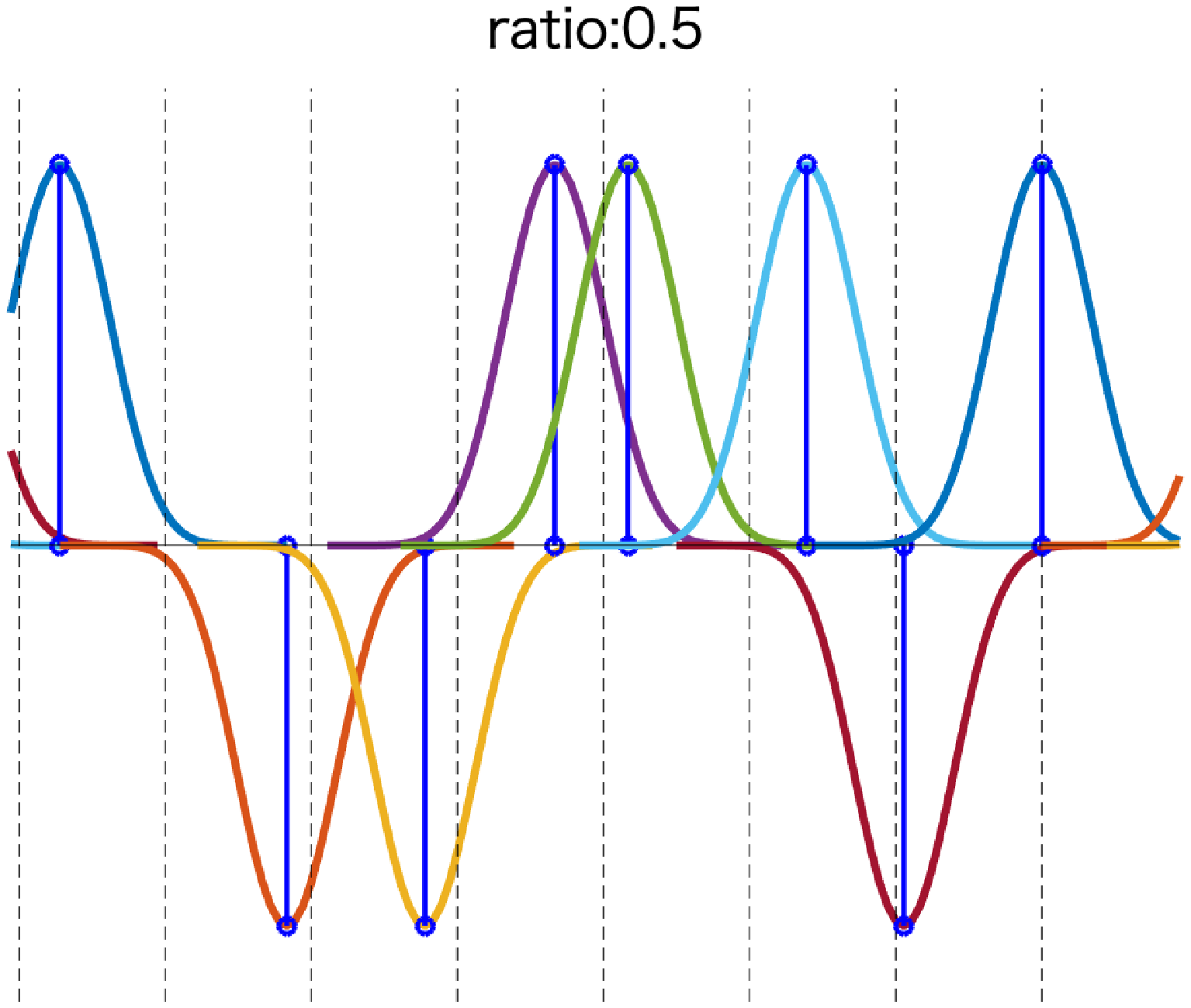}
\hfill
\includegraphics[width=0.48\hsize]{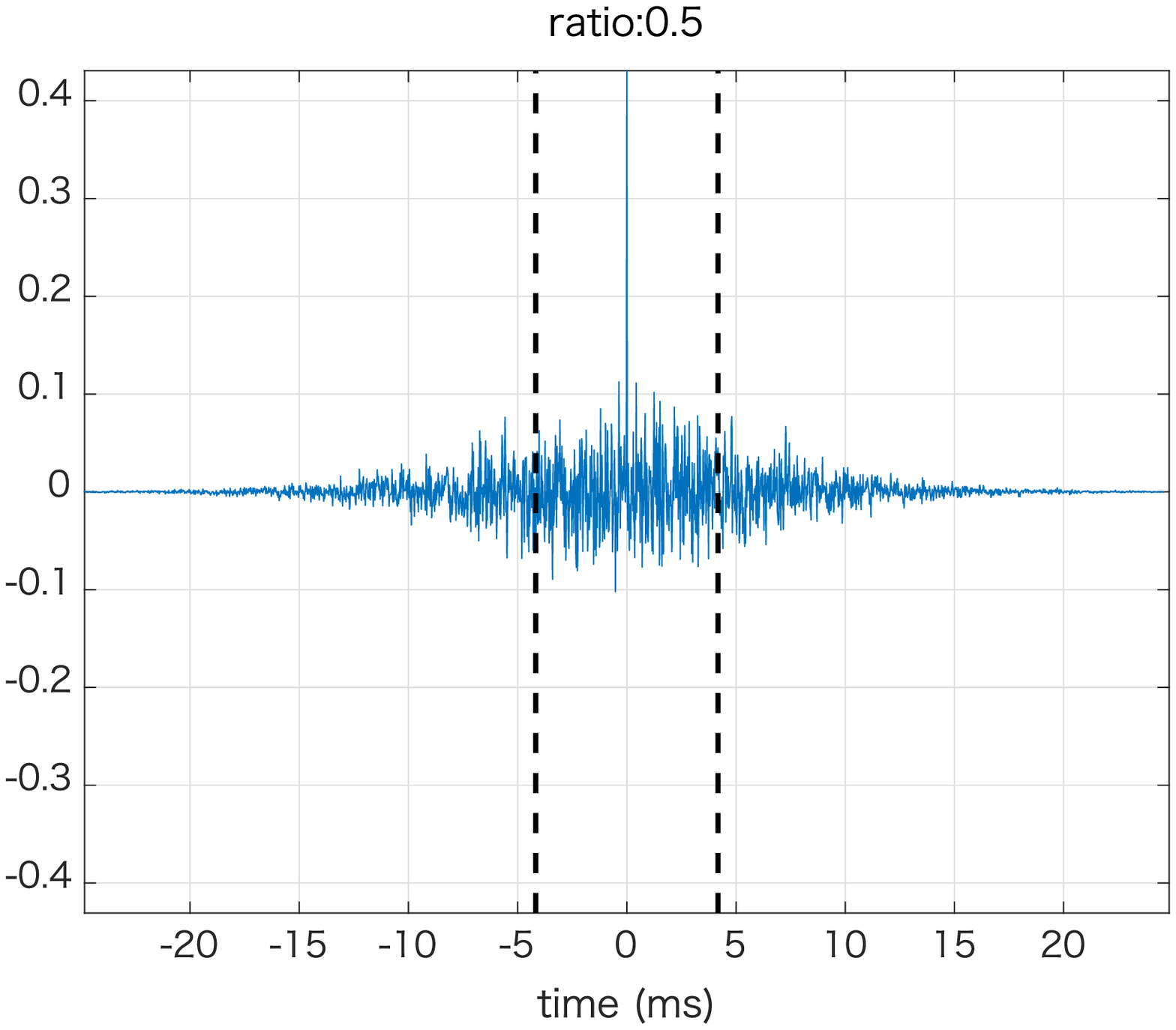}
\includegraphics[width=0.48\hsize]{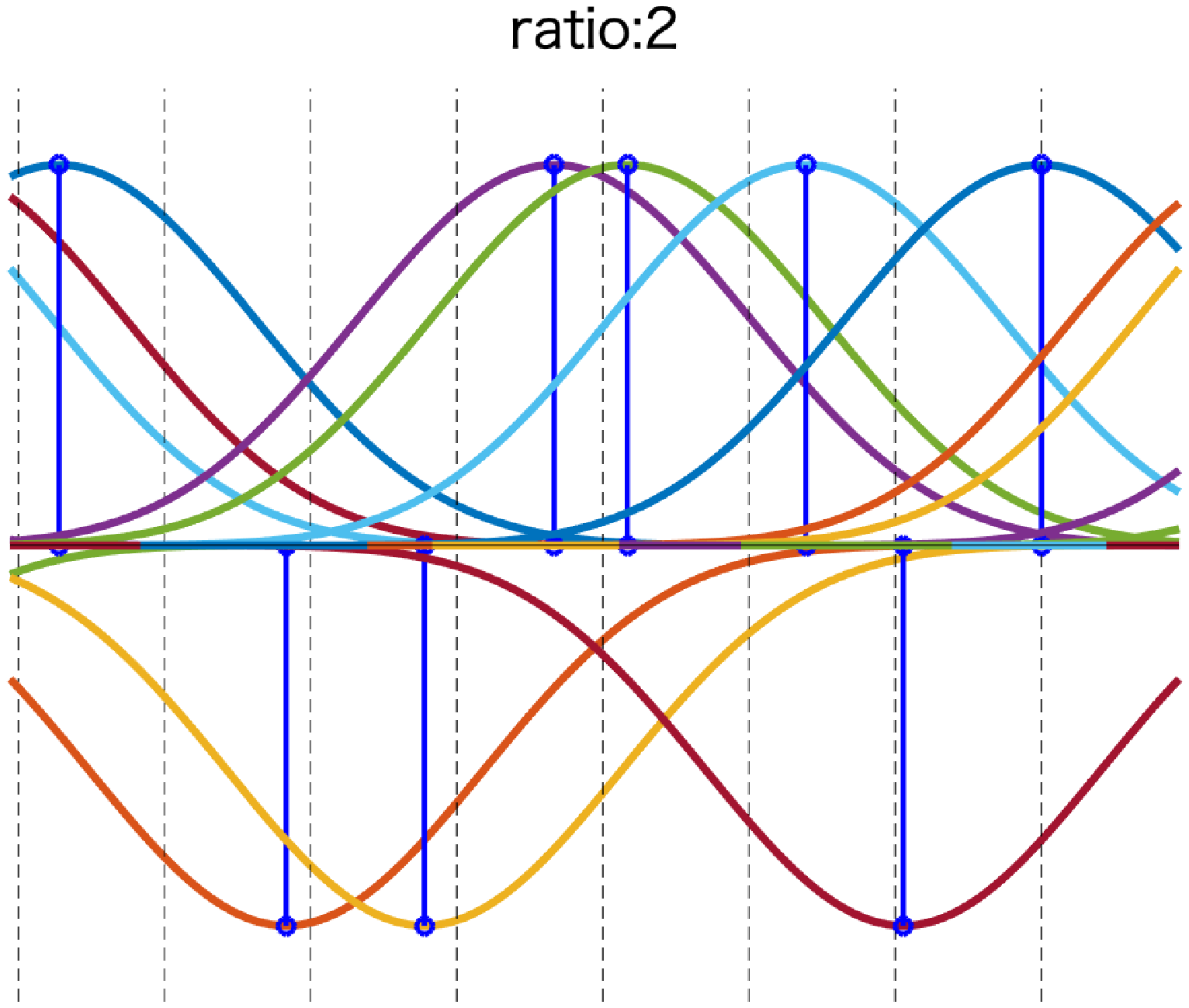}
\hfill
\includegraphics[width=0.48\hsize]{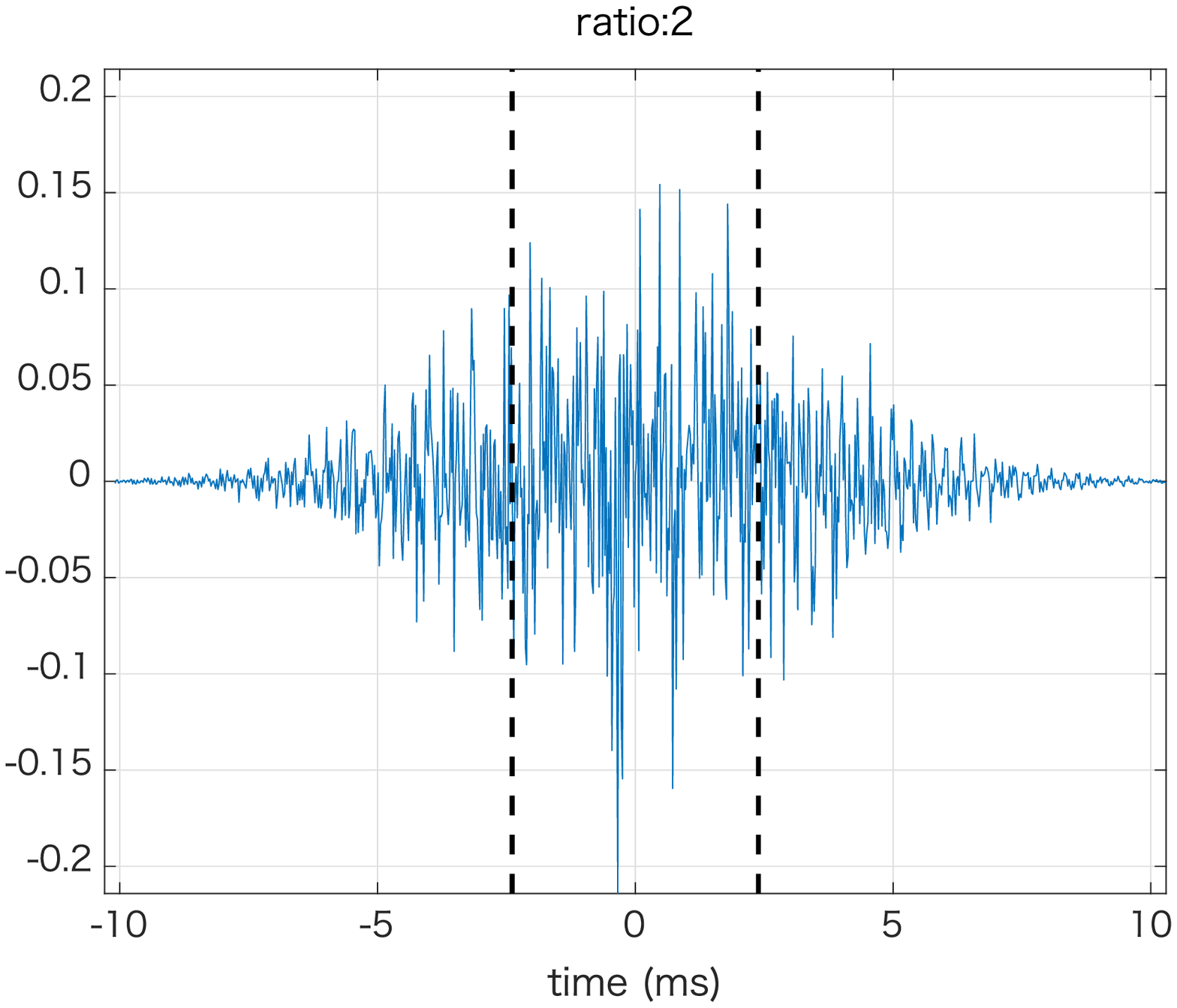}
\caption{(Left plots) Example allocation of phase manipulation functions.
Dotted vertical lines represent the boundaries of the frequency division of
each phase manipulation. Blue solid lines represent the
location of each phase manipulation.
The smooth shapes with color represent allocated phase manipulation functions.
(Right plots) Examples of the generated FVN waveform using the phase manipulations
shown in the left-side plots.
The vertical dashed lines represent $\pm \sigma_{T}$.}
\label{fvnAlloc1}
\end{center}
\end{figure}
Figure~\ref{fvnAlloc1} shows the phase manipulation in the frequency domain
and the corresponding FVN waveforms in the time domain.
The shape of the envelope of FVN waveform depends on the
ratio between the width of the unit phase manipulation function and the
average frequency distance of each manipulation.
Based on a set of simulation tests (refer to Appendix~\ref{ss:fvndesign}), we set the design parameters
of an FVN as follows, where $\sigma_T$ represents the
signal duration.
The average frequency separation $F_d = \frac{1}{5\sigma_T}$,
the frequency spread parameter $B_w = 2 F_d$, and the
maximum phase deviation $\varphi_\mathrm{max} = \pi/4$.
This parameter setting makes the temporal envelope of FVN close to Gaussian.

\section{Application to acoustic measurement}
A unit FVN signal $h_\mathrm{fvn}(n)$ is the impulse response of an all-pass filter
represented as $H_\mathrm{fvn}(k)$.
The frequency domain representation of the time-reversed version of an FVN signal
is the complex conjugate of the original FVN, $H_\mathrm{fvn}^{\ast}(k)$.
The convolution of the original FVN and its time-reversed version yields a unit impulse.
Therefore, an FVN is a member of the time-stretched pulse (TSP\cite{aoshima1981jasa}) in the broadest sense.

Similar to other TSP signals (such as SS and MLS), FVN signals are useful for measuring acoustic impulse responses
because measurement of acoustic systems requires the test signals to reside inside the appropriate operation level.
FVNs have additional design flexibility because they consist of multiple phase-manipulation functions.
This flexibility enables the simultaneous measurement of
multiple acoustic impulse responses.
The flexibility also enables simultaneous measurement of
an impulse response (representing the time-invariant linear system behavior)
and the nonlinear component of an acoustic system.

In addition to these, it is a common practice to shape the long-time averaged power spectrum of the test signal,
to make the signal to noise ratio of each frequency band similar.
For stretched sinusoids, a class of TSP, time axis warping provides such spectrum shaping\cite{morise2007warped,ochiai2011impulse}.
For FVNs, filtering enables such spectrum shaping.
Using an IIR filter for shaping and using the corresponding FIR filter is a useful implementation,
especially for FVNs.

\subsection{Configuration of measurements}

\begin{figure}[tbp]
\begin{flushright}
{  }\hfill\includegraphics[width=0.9\hsize]{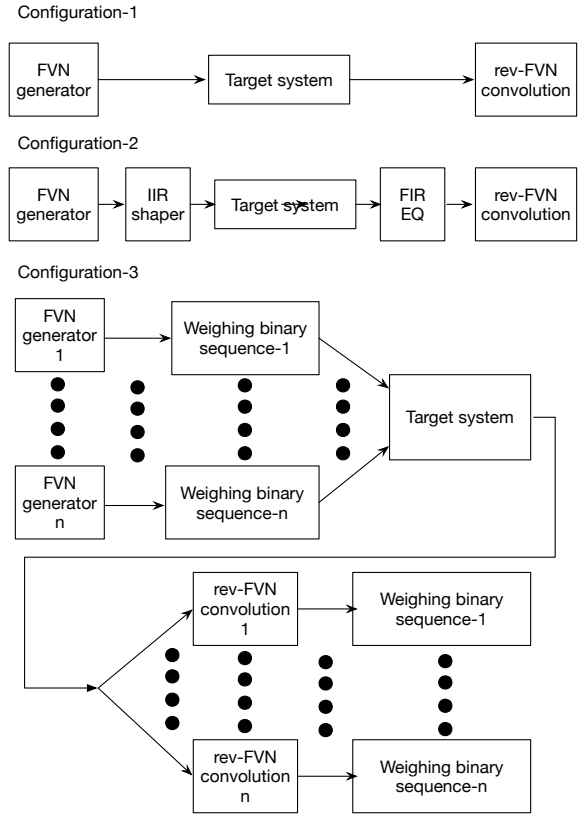}
\caption{Measurement configurations.}
\label{measurngConfig}
\end{flushright}
\end{figure}
Figure~\ref{measurngConfig} shows the typical measurement configurations.
Descriptions of each configuration follow.

\subsubsection{Configuration-1: baseline}
The first configuration-1 directly use an FVN for the test signal for the target system.
The output signal is the convolution of the test signal and the impulse response of the target system.
By convolving with the time-reversed version of the FVN used for the test-signal
makes the equivalent input signal a unit pulse.
Therefore, as far as the target system is linear time-invariant,
the final processed signal represents the impulse response of the target system.

\subsubsection{Configuration-2: spectral shaping}
In usual conditions for measuring acoustic impulse response,
the background noise has higher energy in the low-frequency region.
It makes measurement results in the low-frequency region less reliable.
Shaping the long-term spectrum of the test signal to behave similarly to the
background noise solves this problem.
Measurement results using the shaped test signal
makes each frequency band have similar SNR\cite{morise2007warped}.

For FVN test signals, an IIR (infinite impulse response) filter consisting of
coefficients $\{a_k \}_{k=1}^p$ is adequate for shaping the power spectrum of the test signal
because the same coefficients provide the coefficients of the inverse FIR filter.

Spectral shaping is also a common practice in SS and MLS-based methods\cite{farina2000simultaneous,muller2001transfer,guidorzi2015impulse}.
Many of implementation of spectral shaping in SS-based methods use time-axis warping.
It enables the instantaneous power of the test signal constant.
However, for practical use of acoustic systems, input signals
consist of multiple frequency components and level variations.
This discrepancy between the SS and MLS-based test signals and the actual input signals of acoustic systems
motivated our application of FVNs to nonlinearity measurements\cite{farina2000simultaneous}.

\subsubsection{Configuration-3: simultaneous multichannel measurement}
High degrees of freedom in FVN design enables this configuration.
Assume multiple inputs and a single output system for the target system.
The following procedure provides the set of test signals for this measurement.
1) Prepare different FVN for each input.
2) Prepare a set of binary sequence for each FVN.
The binary sequences are orthogonal each other.
3) For each input, place the prepared FVN repeatedly by determining 
the polarity of the FVN using the corresponding binary sequence. 

The output of the target system is the sum of each response, which corresponds to each FVN.
Convolution with a time-reversed FVN and synchronized averaging
using the corresponding binary sequence yields the impulse response to the selected input.
It is because each response is the sum of the polarity-modulated FVNs.

Preparing this configuration to each output yields
a configuration for measuring MIMO (multi-input and multi-output) systems.

\subsubsection{Configuration-3: nonlinearity measurement}
The configuration-3 is also applicable for nonlinearity measurement.
Assume that the target system is a one-input and one-output system, and
the input is the sum of modulated FVN sequences.
When the target system is a linear time-invariant system,
the calculated impulse response for each FVN is identical.
However, when the target system consisting of nonlinearity,
the processed impulse response deviates from the impulse response of the
linear component.
Because the high degrees of freedom in FVN design makes
different FVNs close to orthogonal each other,
the averaged impulse response derived using different FVN
converges to the impulse response of the linear component.
Therefore the difference between each impulses response and the
average impulse response represents the nonlinear component.

\section{Measurement examples}
We implemented these measurement configurations using MATLAB.
This section introduces measurement examples of each configuration
using several loudspeaker systems and microphones.

\subsection{Configuration-1 and 2: baseline and shaping}

\begin{figure}[tbp]
\begin{center}
\includegraphics[width=0.48\hsize]{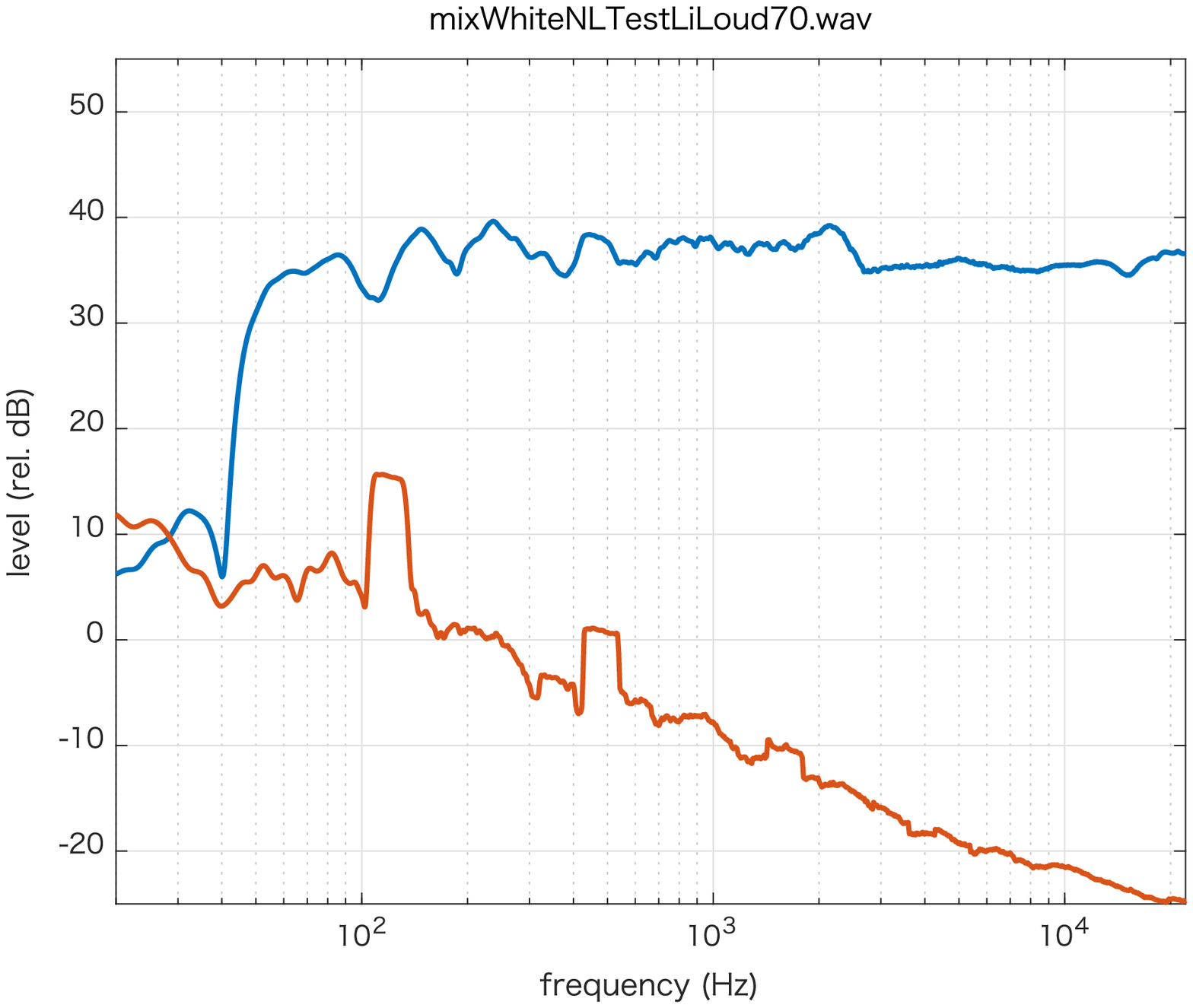}
\hfill
\includegraphics[width=0.48\hsize]{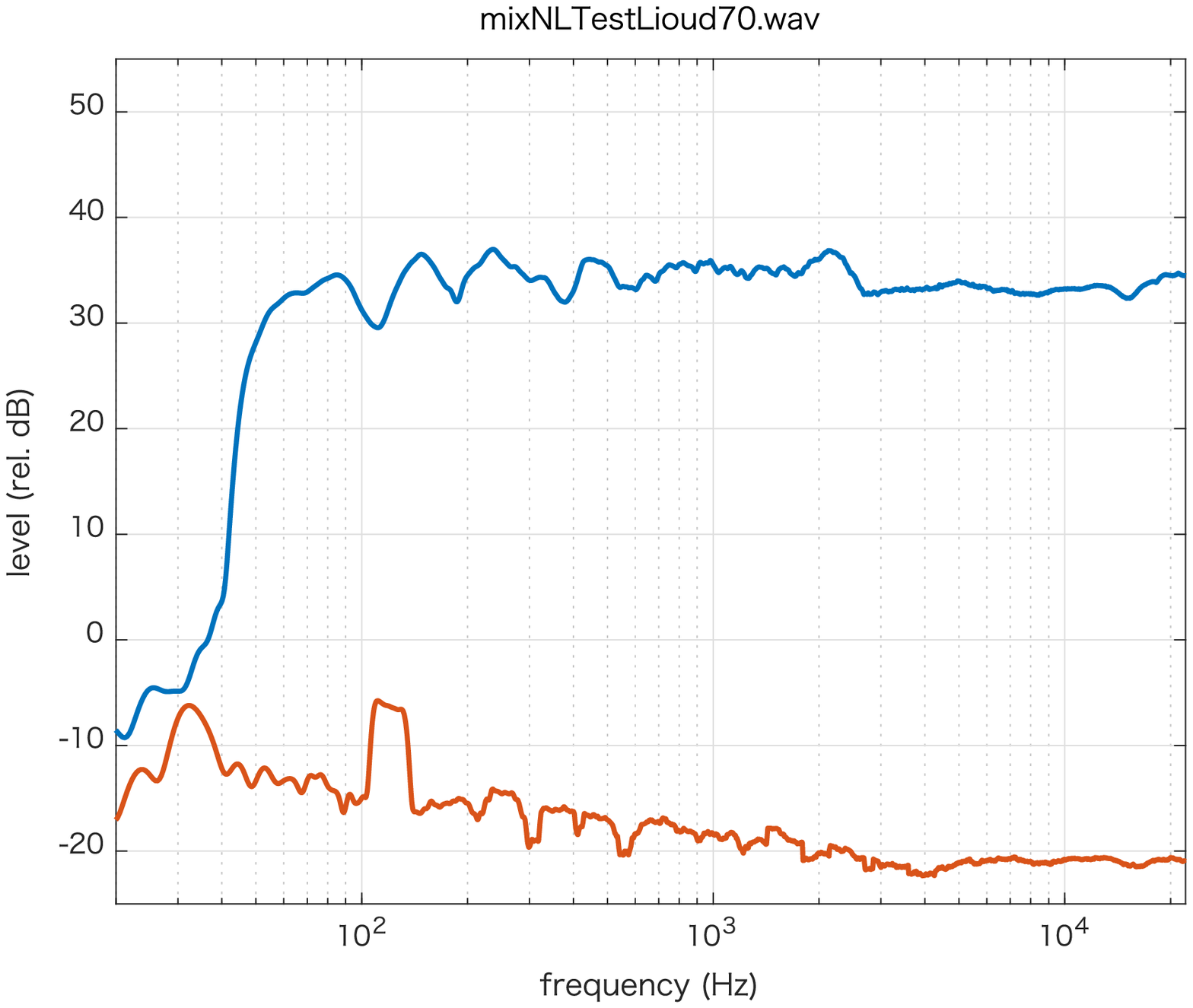}
\caption{Frequency response and the effect of background noise.
The left plot shows the results using configuration-1, and
the right plot shows the results using configuration-2 with
-3~dB/oct frequency shaping.}
\label{config1}
\end{center}
\end{figure}
Figure~\ref{config1} shows the calculated frequency response and the
effect of the background noise using the configuration-1 and 2.
The tested acoustic system is a compact powered monitor (IK-Multimedia iLoud Micro Monitor).
For capturing the sound, we used a miniature headset microphone (Shure MX153T/O-TQG, omnidirectional).
The distance from the microphone to the tweeter, which is directly facing the microphone, of the system was 50~cm.
An audio interface (M-AUDIO M-TRACK 2X2M) with 24~bit 44,100~Hz sampling connected these devices to
a computer (MacBookPro 13" 2.7GHz Intel Corei7 with 16GB memory).
The unit FVN uses $\sigma_T = 0.1$ seconds for generation and repeated 448 times at 5~Hz to calculate the
response and the background noise effect using synchronized averaging.
The sound pressure level at the microphone position was 70~dB using A-weighting.
We conducted the measurements in a room of a house in quiet suburbs, and the
background noise levels (A-weighting) were around 37~dB.

These frequency characteristics are smoothed power spectra using the one-third octave rectangular smoother in the frequency domain.
(Please refer to Appendix~\ref{ss:onethirdsms} details of this smoothing and underlying principles.)
Note that the lower end of the frequency response shows saturation caused by the higher noise level.
The spectral shaping of the test signal solves this problem.
A 46-tap IIR filter shaped the test signal in configuration-2.

\subsection{Configuration-3: nonlinearity measurement}

\begin{figure}[tbp]
\begin{center}
\includegraphics[width=0.9\hsize]{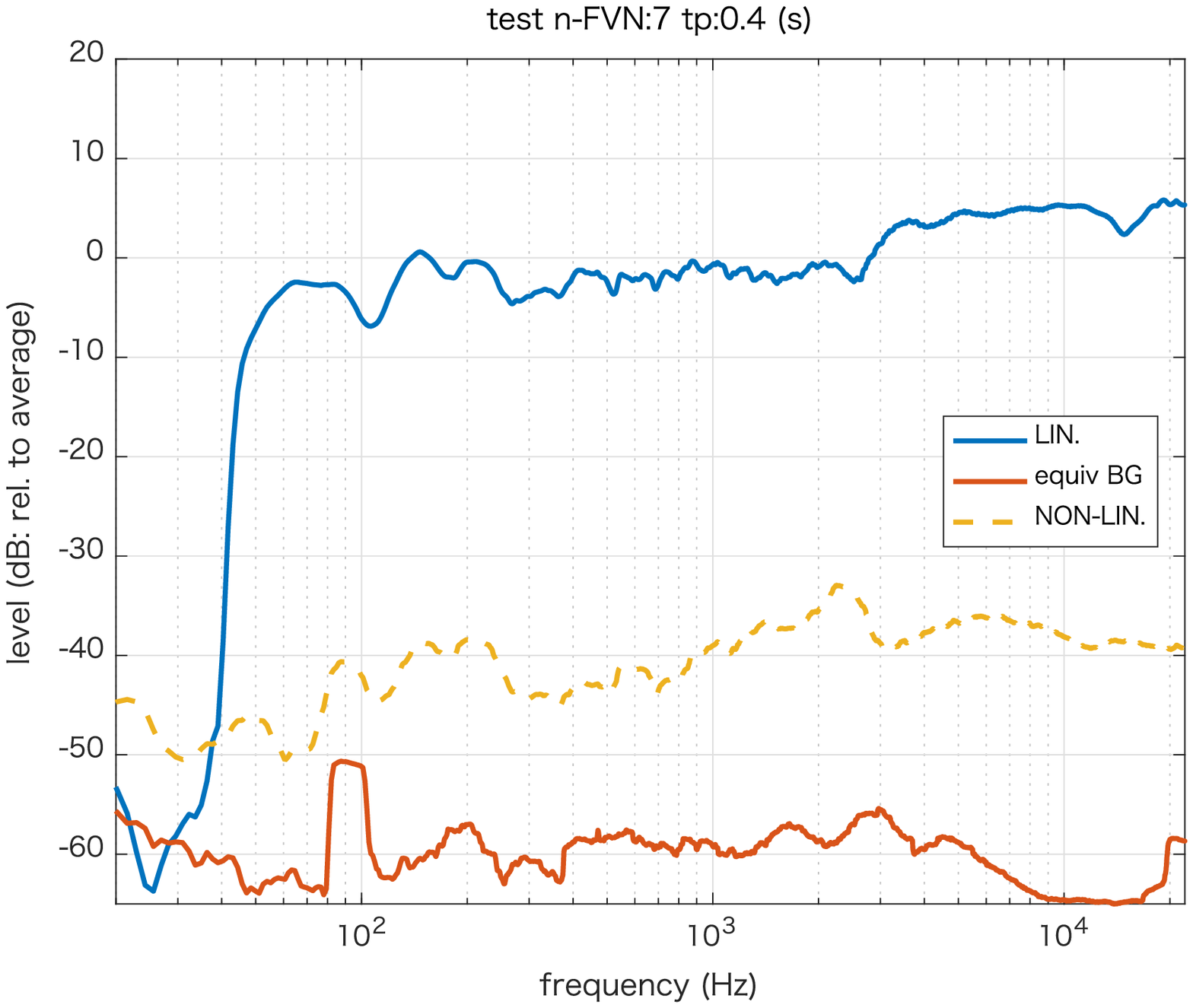} \\
\caption{Measured frequency gain, background noise, and
non-linear components of a compact powered monitor loudspeaker system.
Seven different FVN sequences are mixed using a set of orthogonal binary weights.}
\label{simlNL1}
\end{center}
\end{figure}
Figure~\ref{simlNL1} shows a result of configuration-3 applied to nonlinearity measurement.
In this example, we generated seven different FVNs
and modulated polarity using binary sequences orthogonal each other, for example,
$[1, 1, 1, 1]$ and $[1, -1, 1, -1]$.
Refer Appendix~\ref{ss:sequencedesign} for the details of the test signal design.
The sound pressure level was 80~dB in this test.
In this experiment, we used the other omnidirectional miniature condenser microphone (DPA-4066B). 
The other test conditions are the same to configuration-1 and 2.
The length of the test signal is 108~s.
We also recorded background noise in the same length for evaluating effects on the resulted response.
The total time for this measurement was 245~s.
This condition effectively averages 896 (128 repetitions of unit FVNs times seven mixing sequences) individual impulse responses
and the filtered (by FVNs) background noise.

The blue line shows the averaged frequency response representing the linear component.
The red line shows the averaged effect of background noise.
For representing the response due to nonlinearity, which is the primary interest,
we used a dashed yellow line.

\begin{figure}[tbp]
\begin{center}
\includegraphics[width=0.9\hsize]{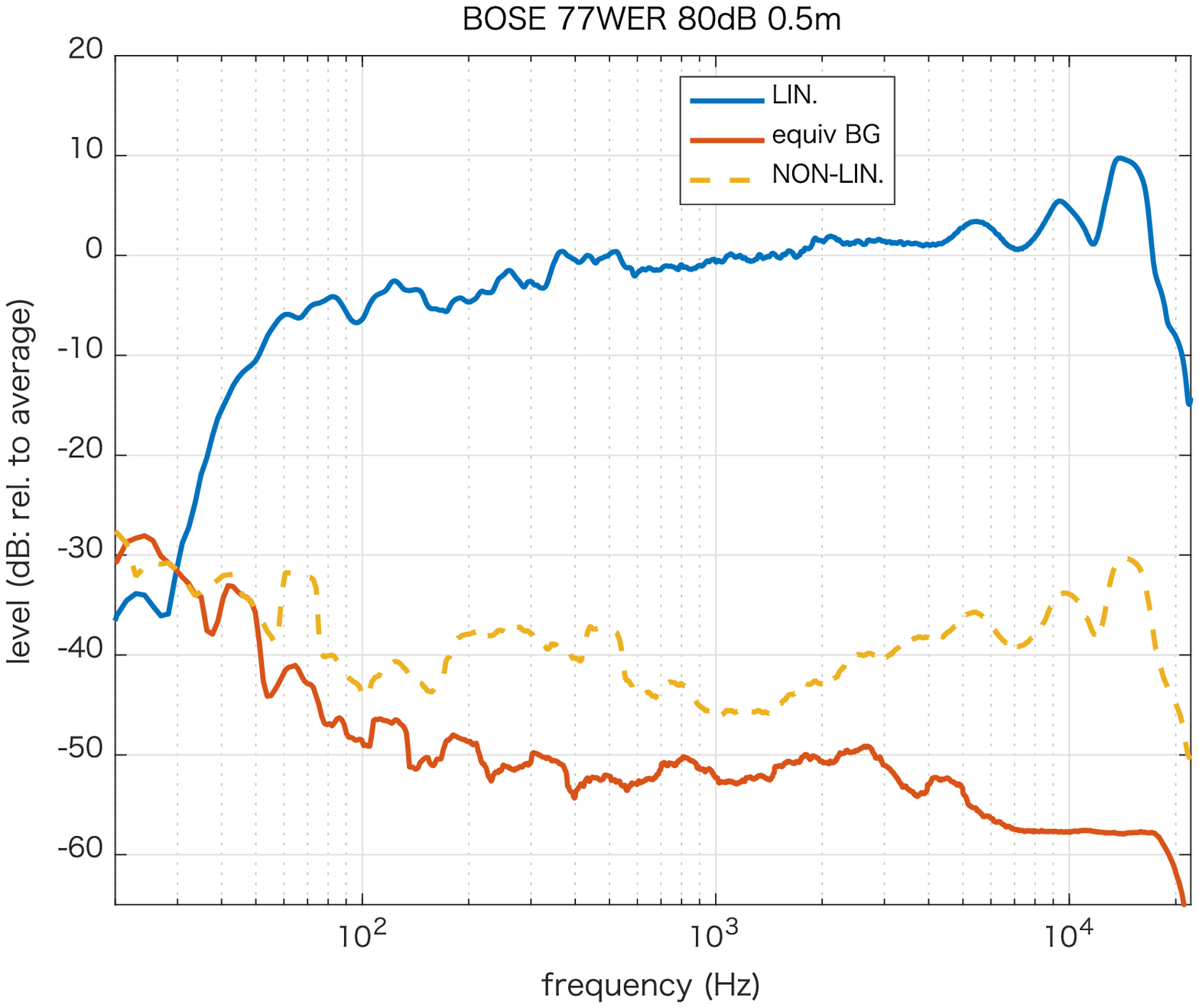} \\
\vspace{2mm}
\includegraphics[width=0.9\hsize]{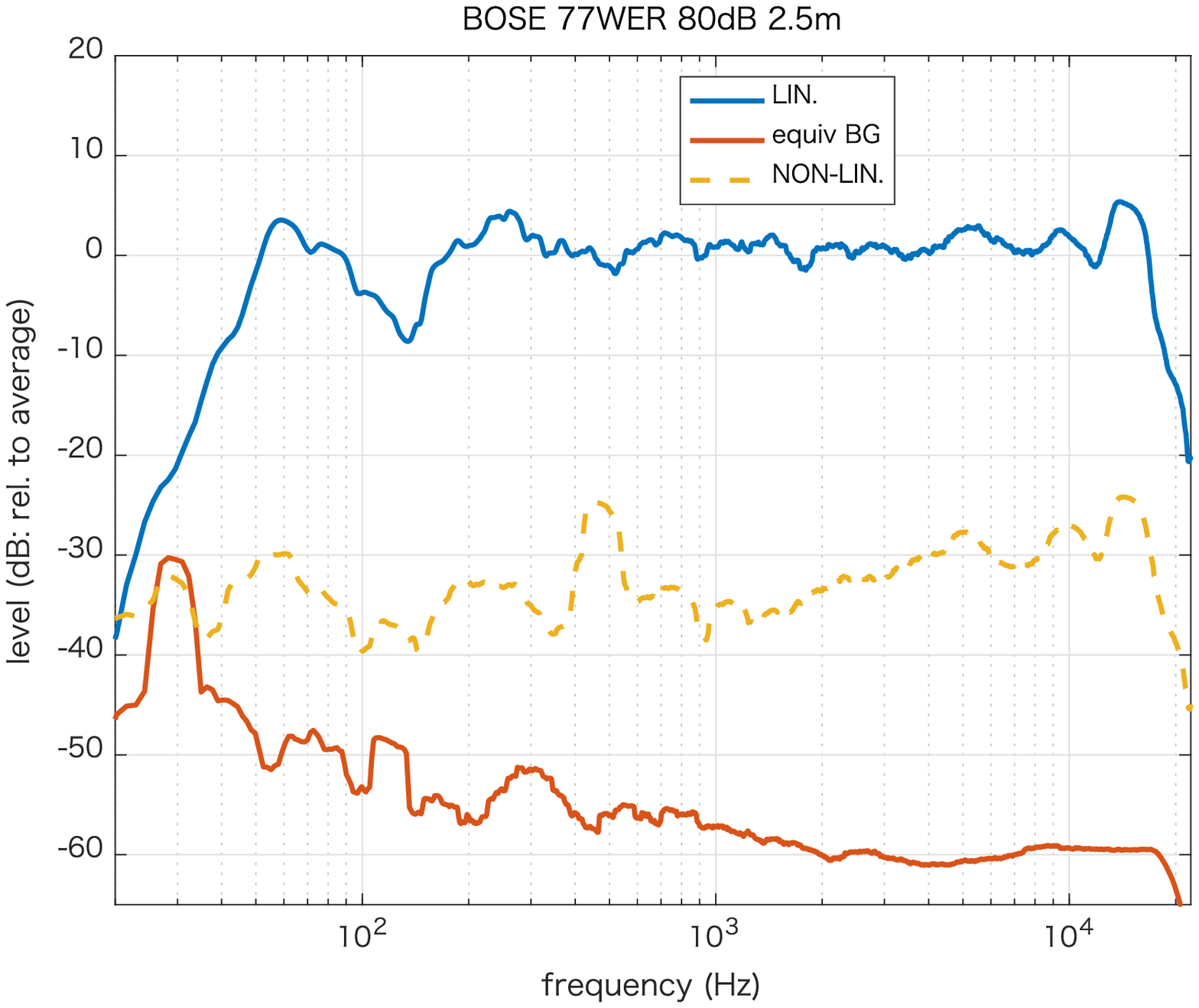} \\
\caption{Measured frequency gain, background noise, and
non-linear components of a tall-boy type passive loudspeaker system (BOSE 77WER).
Four different FVN sequences are mixed using a set of orthogonal binary weights.
The sound pressure level at the microphone was 80~dB.
The upper plot shows the results at 50~cm distance, and
the lower plot shows the results at 2.5~m.}
\label{fvn44k20190628T160125}
\end{center}
\end{figure}
\begin{figure}[tbp]
\begin{center}
\includegraphics[width=0.9\hsize]{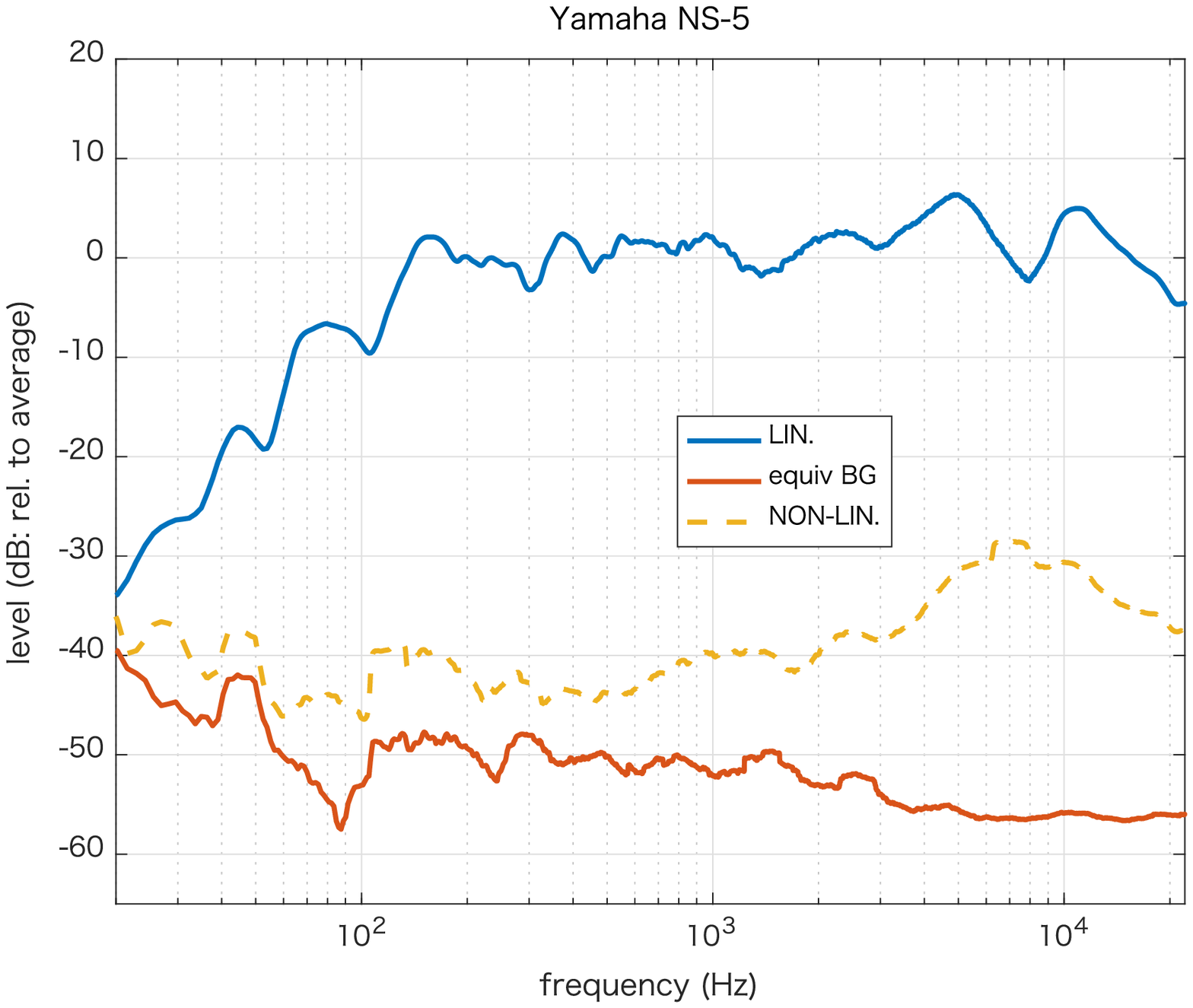} \\
\vspace{2mm}
\includegraphics[width=0.9\hsize]{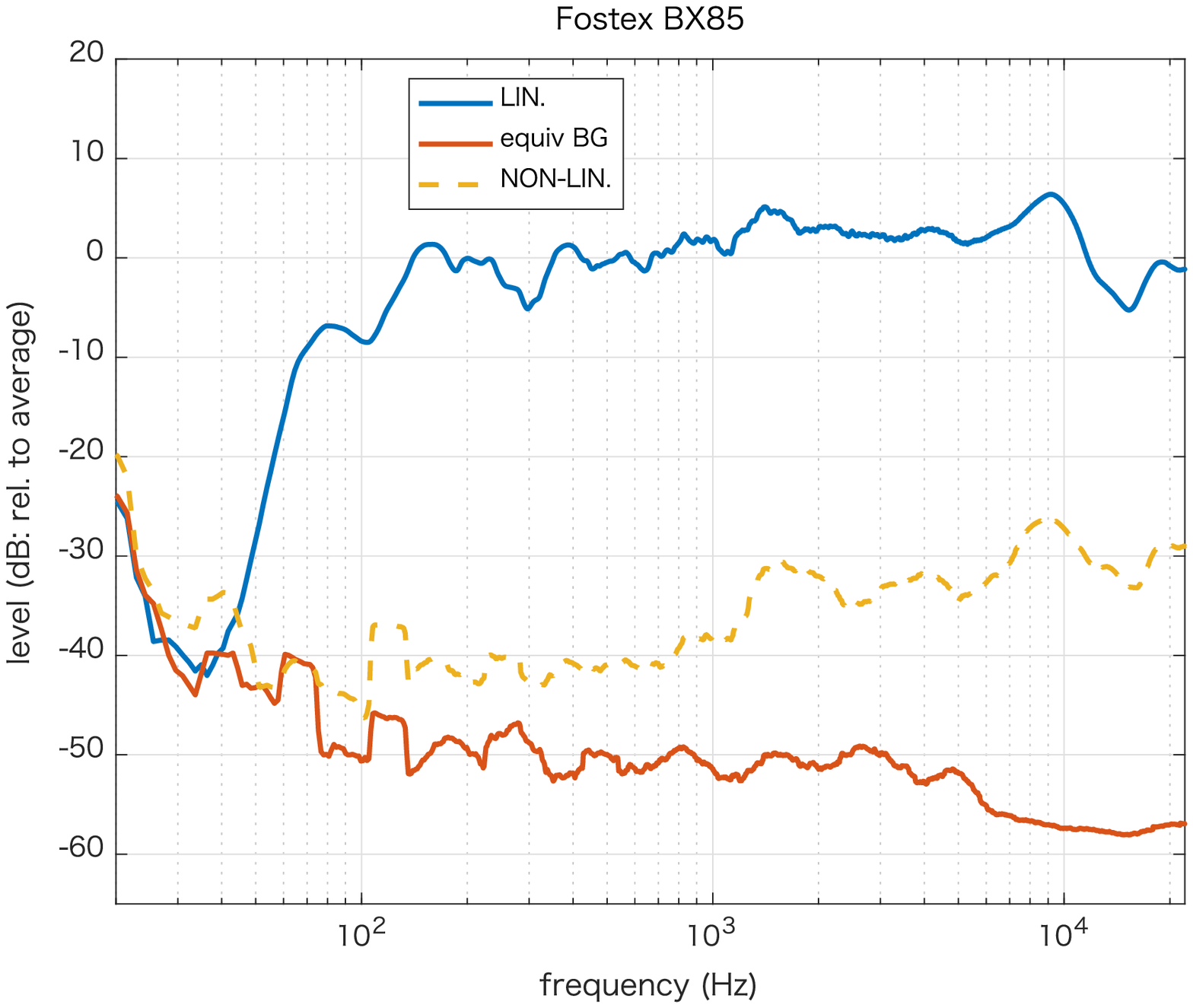} \\
\caption{Measured frequency gain, background noise, and
non-linear components of two compact passive speakers.
Four different FVN sequences are mixed using a set of orthogonal binary weights.
The upper plot shows the result of a two-way system (Yamaha NS-5)
and the lower plot shows the result of a full-range system (Fostex FF85WK unit in BK85WB2 box).
The distance to the microphone is 50~cm.}
\label{fvn44k20190628T150656}
\end{center}
\end{figure}
Figures~\ref{fvn44k20190628T160125} and \ref{fvn44k20190628T150656} show results
using a simplified procedure.
It uses four FVN mixtures.
It requires 30~s to measure, including nonlinearity and background noise measurement.
Note that the room used for compact system measurements has resonances caused by standing waves.
They are 45, 74, 86, and 160~Hz.
Effects of higher resonances are not visible because of the one-third octave smoothing
explained in Appendix~\ref{ss:onethirdsms}.

We prepared a MATLAB function which implements these measurement procedures and
the calibration procedure of the microphone sensitivity.
This function also provides a quick measurement mode.
In the quick mode, 
which skips the nonlinearity measurement, the typical measurement time is eight seconds.

\section{Discussion}
In multiple acoustic system measurements, the recorded signal is a single channel data.
(It may also record the monitor signal to the acoustic system's input simultaneously. This setting yields two-channel data.)
The same demultiplexing procedure separates individual responses from the
recorded single-channel data.
Combination of these multiple acoustic system measurements and the previously introduced nonlinearity measurements
is also possible.
However, in practice, measuring linear components only is useful for implementing interactive and real-time measuring tools.
Appendix~\ref{ss:rtsttool} briefly introduces an example implementation.

The actual acoustic systems consist of temporal variations.
For example, airflow such as breeze or wind modulate the speed of sound
and introduces warping of the time axis.
The difference of the master clock of the output and the input devises
introduce drift of the time axis.
These distort the calculated impulse responses.
There are several ways to estimate the warping of the time axis using the test signal itself.
Applying the inverse function to compensate for the time axis warping before
the pulse recovery procedure, it yields appropriate impulse responses.
The test signals based on FVN is less sensitive to these errors
than MLS-based methods and SS-based methods.
This tolerance to this time axis warping of FVN is a generalization of ``pure white pseudonoise'' idea\cite{Mori2017ast}.
Systematic investigations of these are topics of further research.

\section{Conclusions}
We introduced a set of applications of the frequency domain variant of velvet noise (FVN)
for acoustic measurements.
Specifically, the efficient simultaneous measurement of the linear and the nonlinear component of the
acoustic system,
and simultaneous measurement of multiple acoustic systems are the significant contributions of this article.
The introduced tools and the other assistive tools, SparkNG are available in
the first author's GitHub repository\cite{kawaharaGit}.

\section*{Acknowledgment}

The authors wish to thank Yutaka Kaneda, professor at Tokyo Denki University, for discussions on Swept-Sine and MLS signals.
KAKENHI  (Grant in Aid for Scientific Research by JSPS) 16H01734, 15H03207, 18K00147, and 19K21618 supported this research. JST PRESTO Grant Number JPMJPR18J8 also supported it. 

\bibliographystyle{IEEEtran}

\bibliography{APSIPA19FVNtools.bib}

\renewcommand\thefigure{\thesection.\arabic{figure}} 
\appendix
\setcounter{figure}{0}

\subsection{Simulations for FVN design}\label{ss:fvndesign}
We conducted a set of simulations to design a unit FVN as the building block of the test signals.
The sampling frequency is 44,100~Hz in the following simulations.
We also conducted simulations for
constructing test signals by
allocating unit FVNs on the time axis.
The allocation uses a set of mutually orthogonal binary sequences 
for measurement configuration-3.

\subsubsection{Unit FVN design}

\begin{figure}[tbp]
\begin{center}
\includegraphics[width=0.9\hsize]{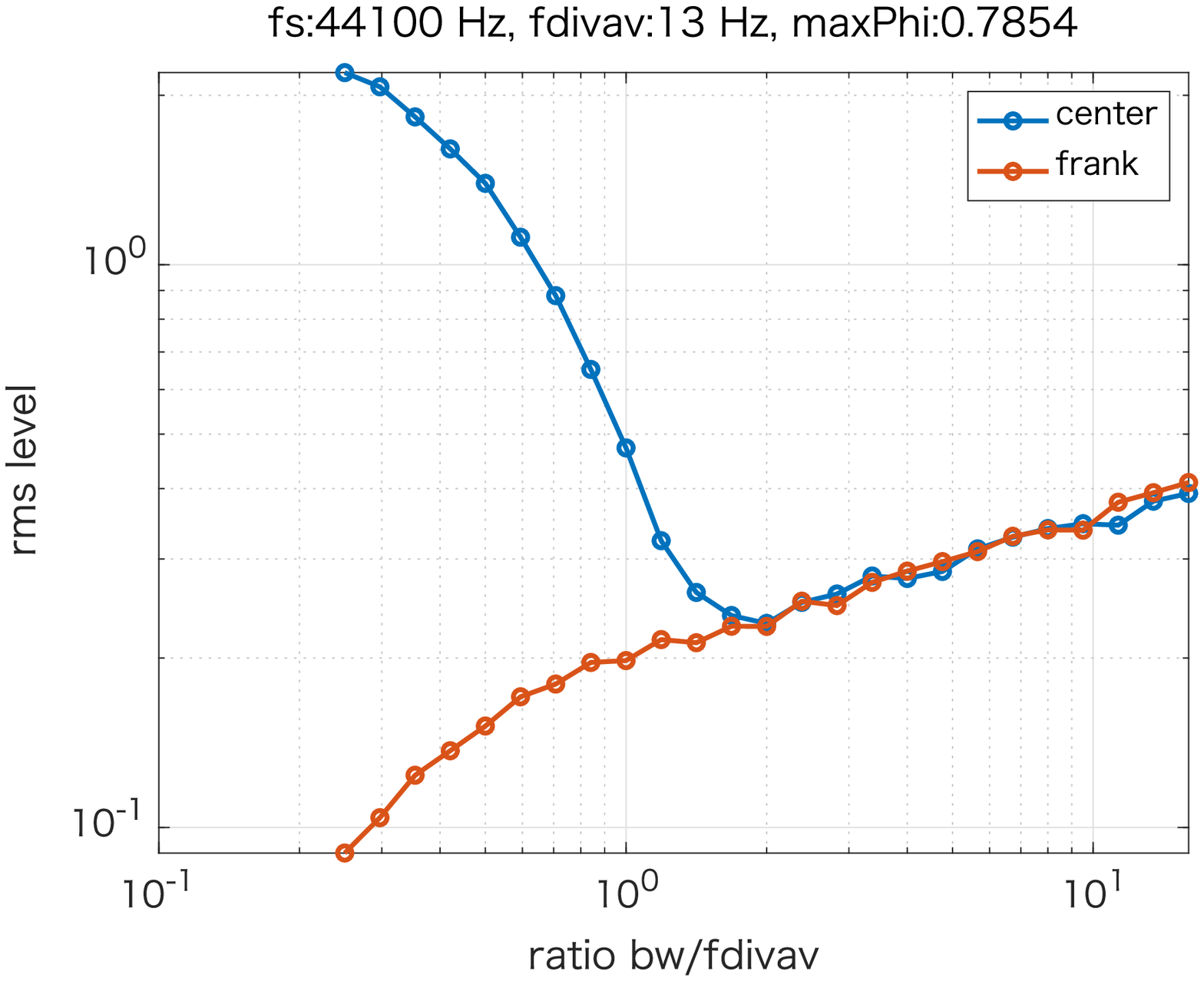} \\
\hfill
\vspace{1mm}
\includegraphics[width=0.9\hsize]{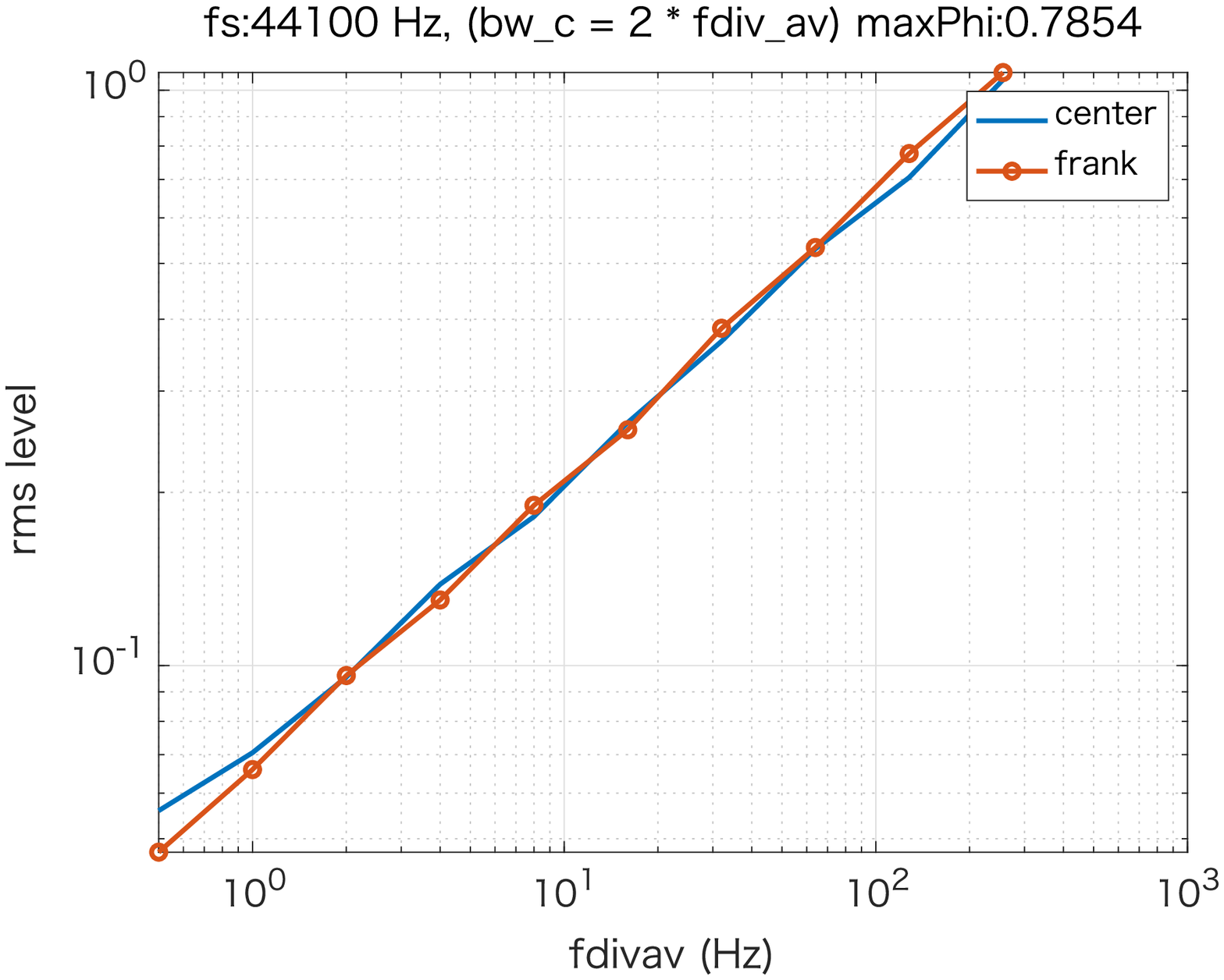} \\
\hfill
\vspace{1mm}
\includegraphics[width=0.9\hsize]{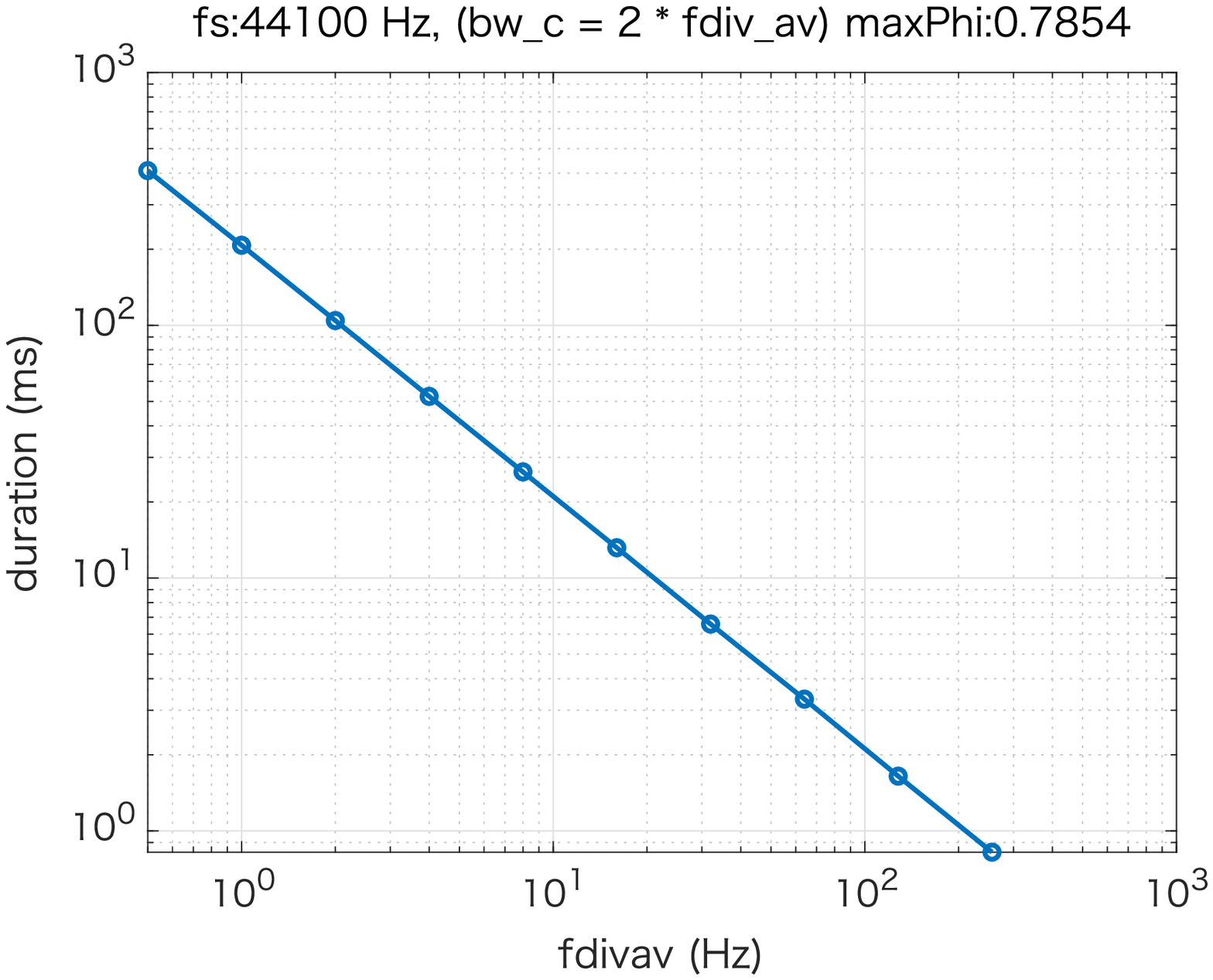} \\
\caption{(Top) RMS values of center 9 points and
the franking 10 points as functions of the ratio.
(Middle) RMS values of center 9 points and
the franking 10 points as functions of average frequency interval $F_d$.
(Bottom) Duration plot as a function of average frequency interval $F_d$.\label{bwfavratioAndShapeGauss}}
\end{center}
\end{figure}
Figure~\ref{bwfavratioAndShapeGauss} shows excerpts from the simulations.
First, we found the ratio $B_w/F_d$ determines the temporal shape of the envelope
for different $B_w$ and $F_d$ combinations.
The top panel of Fig.~\ref{bwfavratioAndShapeGauss} indicates that
the ratio higher than 2 provides a smooth envelope shape.
The middle plot verifies that the ratio is the governing factor.
The bottom plot provides the value of $F_d$ based on the given duration $\sigma_T$.

\subsubsection{Test signal design}\label{ss:sequencedesign}

\begin{figure}[tbp]
\begin{center}
\includegraphics[width=0.85\hsize]{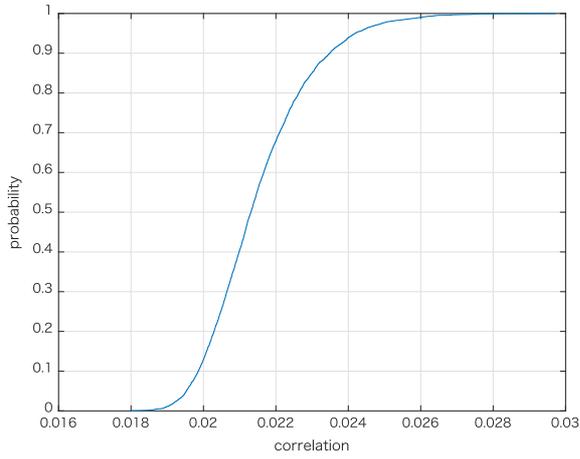} \\
\end{center}
\caption{Distribution of the maximum absolute values of cross-correlations
of unit FVNs. The duration used for designing FVNs was $\sigma_T = 100$~ms.}
\vspace{-1mm}
\label{fig:xcorrfvns100mssg}
\end{figure}
Figure~\ref{fig:xcorrfvns100mssg} 
shows the distribution of the maximum absolute values of
the cross-correlation between different FVNs.
We used the duration of $\sigma_T = 100$~ms for designing each FVNs.
This small cross-correlation indicates that FVNs are mutually close to orthogonal.
However, these cross-correlation prevents mixing FVN sequences to
measure multiple systems simultaneously.

We introduced a set of binary sequences $b_k[n]$ which are orthogonal each other
where $k$ represents the identifier of the sequence and $n$ represents the position
in the sequence.
We used the following sequence.
\begin{align}\label{eq:orthSeq}
\{b_1[n]\}_{n=1}^N & = [1, 1, 1, 1, \ldots, 1, 1] \nonumber \\
\{b_2[n]\}_{n=1}^N & = [1, -1, 1, -1, \ldots, 1, -1] \nonumber \\
 & \vdots \nonumber \\
\{b_k[n]\}_{n=1}^N & = [\overbrace{1, 1, \ldots, 1, 1}^{2^{(k-2)}}, \overbrace{-1, -1, \ldots, -1, -1}^{2^{(k-2)}}, \ldots]\nonumber \\
 & \vdots \nonumber \\
\{b_K[n]\}_{n=1}^N & = [\overbrace{1, \ldots, 1}^{2^{(K-2)}}, \overbrace{-1, \ldots, -1}^{2^{(K-2)}}, \ldots]
\end{align}
The length of the sequence $N = 2^{K+1}$

Let $\mathrm{B}$ a matrix consisting of $\{b_k[n]\}_{n=1}^N$ for each row.
Then, it follows.
\begin{align}
\mathrm{B} \mathrm{B}^T = N \mathrm{I} ,
\end{align}
where $\mathrm{I}$ represents the identity matrix.

The $k$-th FVN sequence places a unit FVN defined by (\ref{eq:unitfvn}) at every
$n_o$ samples on a discrete-time axis using $b_k[n]$ for defining its polarity.
We set the period $n_o$ longer than the effective response length of the target system
to prevent interference caused by circular convolution.

We use $b_k[n]$ for weighting individual impulse responses when calculating the averaged
impulse response.
This arrangement cancels interferences between different FVN sequences and
enables the simultaneous measurement of multiple acoustic systems.
Note that for the synchronized averaging we use the central $2^K$ elemental results.
This selection is necessary for canceling interferences precisely.

For using these sequences to nonlinearity measurement,
we add all these sequences to generate the test signal.
For using these sequences to measure multiple acoustic systems simultaneously,
we feed each sequence to each acoustic system and use
one microphone to acquire the mixed responses.
\begin{figure}[tbp]
\begin{center}
\includegraphics[width=0.89\hsize]{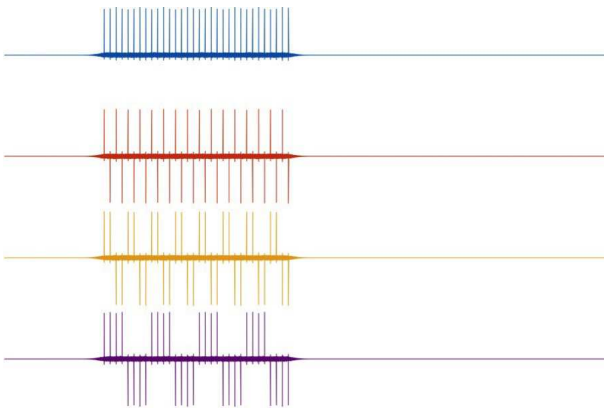}  \\
\end{center}
\vspace{2mm}
\caption{Demultiplexed test signals recorded from
the monitor output of the audio interface.}
\label{fig:refSigDeconv}
\end{figure}

Figure~\ref{fig:refSigDeconv} 
illustrates how each FVN sequence looks.
Each line shows the convolution with the time-reversed corresponding FVN.
The first half provides the impulse response, and the latter half provides the background noise.
Note that the first part of each line shows residual signals between pulse positions.
They are interferences caused by cross-correlation between different FVN sequences.

\subsection{One third octave smoothing and impulse response}\label{ss:onethirdsms}
\begin{figure}[tbp]
\begin{center}
\includegraphics[width=0.95\hsize]{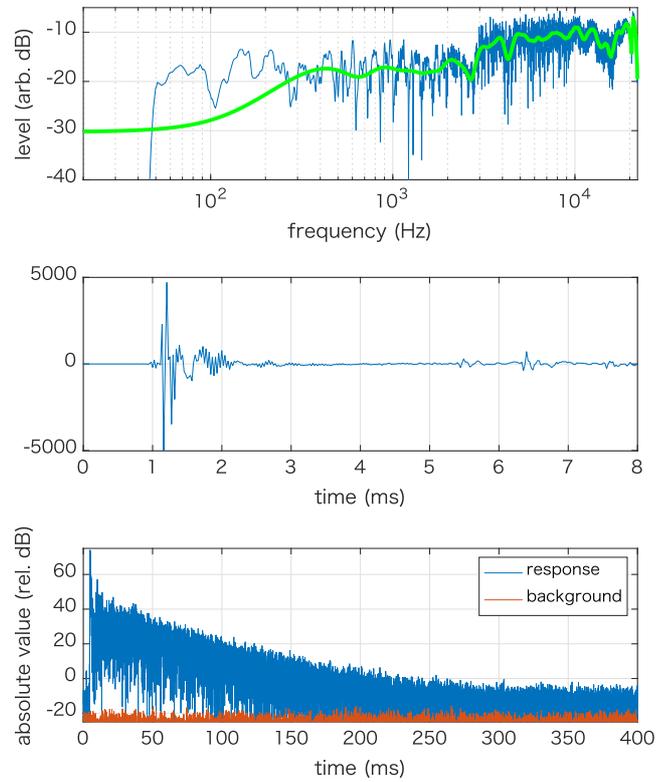} \\
\end{center}
\caption{(Top) Frequency response calculated using 400~ms (blue line) and
the initial 3.2~ms (green line).
(Middle) The initial 8~ms of the averaged impulse response.
This time axis compensates for the propagation time from the loudspeaker to the microphone.
(Bottom) The averaged impulse response and the averaged background noise.}
\label{impRespAndFreqResp}
\end{figure}
Figure~\ref{impRespAndFreqResp} 
shows the raw results of the measurement.
The top panel shows the frequency responses.
The blue line represents the results using the whole 400~ms response waveform,
and the thick green line shows the results using the initial 3.2~ms response,
where the response consists of only the direct path.
The middle plot shows the response waveform of the initial 8~ms.
The response indicates that reflection from the ceiling and the floor are visible from 5 to 7~ms.
The bottom plot shows the whole waveform of the system response and the
filtered background noise.

The frequency response derived from the whole 400~ms response has many sharp peaks and dips.
They are the results of many reflections.
The response waveform which does not consist of any reflections
provide smooth representation.
However, this short response introduces a significant error in the lower frequency region.
This inaccuracy is inevitable.
We introduce a one-third octave smoothing for visualizing the frequency response.

Let $P_w(f)$ represent a power spectral representation of the frequency response.
The following equation provides the smoothed representation $Q(f)$.
\begin{align}
Q(f) & = \frac{1}{f_H - f_L} \int_{f_L}^{f_H} P_w(\nu) d\nu \\
\mbox{where} & \nonumber \\
f_H & = 2^{\frac{1}{6}} f \\
f_L & = 2^{-\frac{1}{6}} f .
\end{align}

\begin{figure}[tbp]
\begin{center}
\includegraphics[width=0.9\hsize]{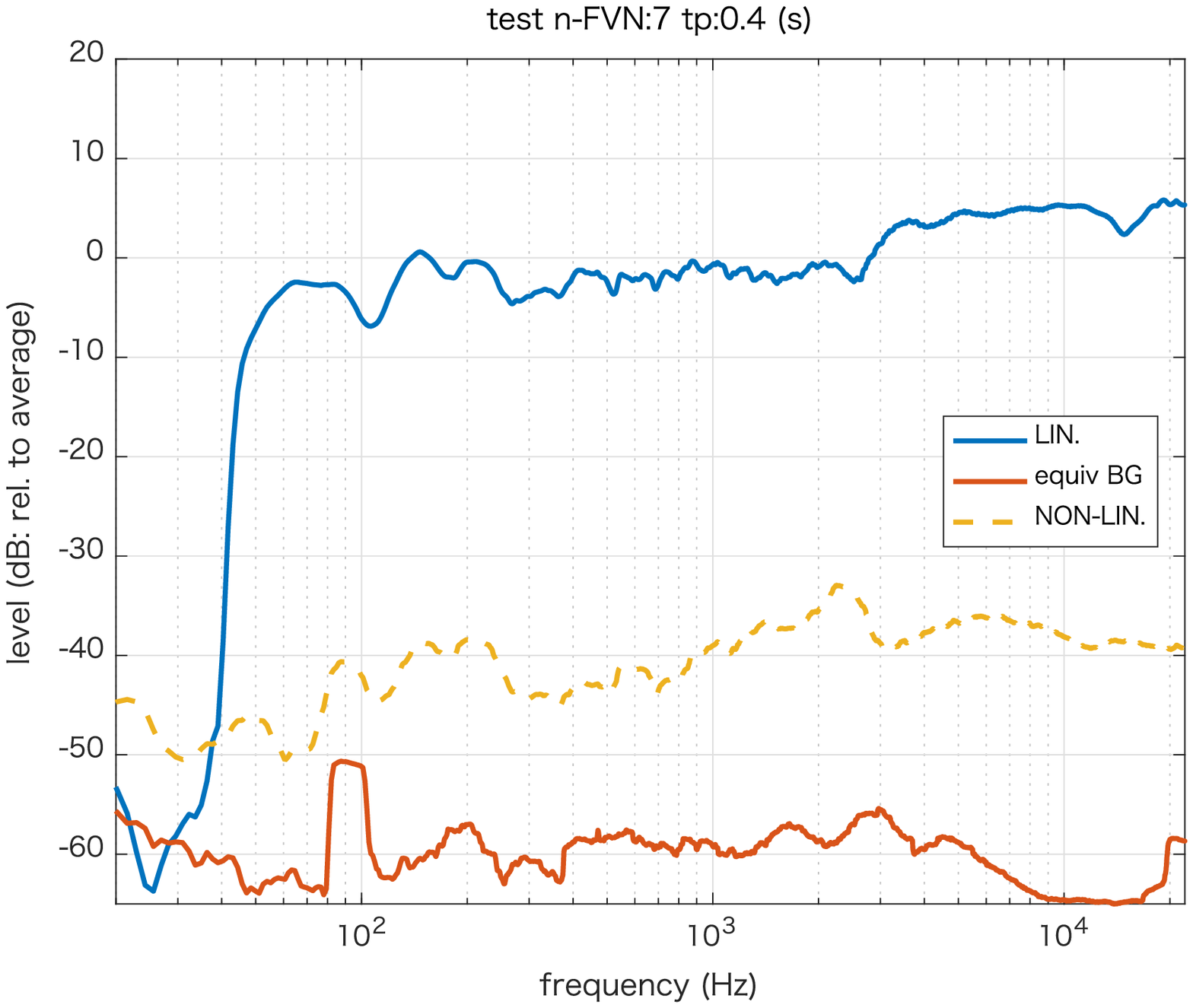} \\
\caption{Frequency response of the linear component (blue line).
Effects of the background noise (red line), and
the component due to nonlinearity (yellow dashed line).\label{fvn44k20190701}}
\end{center}
\end{figure}
Figure~\ref{fvn44k20190701} 
shows the
smoothed responses of the same data.
This figure is the same as Fig.~\ref{simlNL1}.

\subsection{Interactive and real-time tool}\label{ss:rtsttool}
\begin{figure}[tbp]
\begin{center}
\includegraphics[width=0.99\hsize]{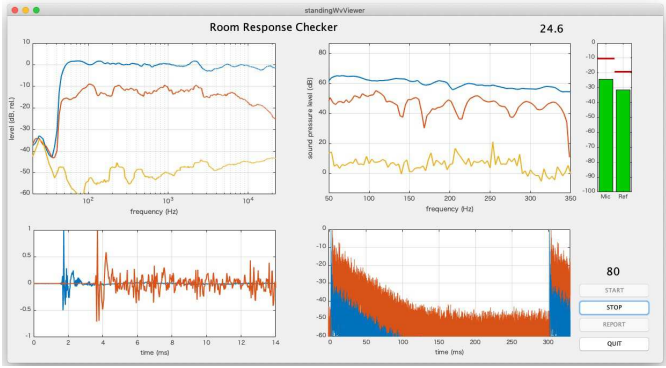} \\
\caption{GUI of an interactive and real-time tool for
two-channel acoustic system measurement.\label{fig:sttoolGUI}}
\end{center}
\end{figure}
Figure~\ref{fig:sttoolGUI} shows a snapshot of the GUI of an interactive and
real-time tool for two-channel acoustic system measurement
using a two-channel FVN-based test signal 
($b_1[n]$ of (\ref{eq:orthSeq}) for the left channel and $b_2[n]$ for the right channel.
Every 0.2~s updates all plots in the GUI.

\begin{figure}[tbp]
\begin{center}
\includegraphics[width=0.99\hsize]{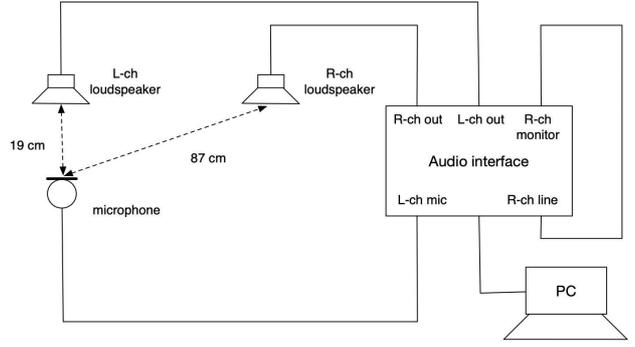} \\
\caption{Schematic diagram of the measurement setting.\label{fig:settingST}}
\end{center}
\end{figure}

Figure~\ref{fig:settingST} shows the setting of the measurement.
The monitor output of the right channel goes to
the channel-2 input.).
The acoustic system consists of two powered loudspeakers (IK Multimedia iLoud Micro Monitor).
The distance from the left channel to the microphone (DPA 4066 miniature omnidirectional condenser microphone)
is 19~cm and that for the right channel is 87~cm.

The top left plot of the GUI shows the amplitude responses 
(blue line: the left channel, red line: the right channel,
and yellow line: measurement error).
This plot uses the same one-third octave smoothing.
The bottom left plot shows the impulse responses showing
the initial 14~ms.
The origin of the time axis is the input pulse location.
(Note that the loudspeaker system consists of a digital
signal processing component which introduces around 1~ms processing delay).
The maximum absolute value of each response is normalized to 1.

The top right plot shows the amplitude responses without smoothing.
The plot shows from 50~Hz to 350~Hz using the linear frequency axis.
The vertical axis represents the calibrated sound pressure level (in dB).
It uses the same color-channel coding with the top left plot.
In this plot, the red line has sharp dips.
These dips represent the effect of standing waves of the room.
While slowly moving the position of the microphone,
the lines in the plot change their shapes dynamically, demonstrating
significant effects of standing waves in the low-frequency region.

The bottom right plot shows the whole impulse responses (absolute values of the response).
The vertical axis is the absolute value of the response in relative dB (maximum response is 0~dB).
The bars of the top right corner shows the input level monitor
where 0~dB corresponds to the maximum input level.
The green bars represent the RMS values, and the red horizontal lines represent
the instantaneous peak values.

The unit length of the component orthogonal FVN sequence of the tool is about 7~s.
It means that the microphone has to stop moving for longer than 7~s to get
a reliable measurement result, while users can move the microphone anytime.
The tool also has calibration and report generation functions.
Our GitHub repository\cite{kawaharaGit} has detailed technical documentation of this tool.

\renewcommand{\textheight}{186mm}

\subsection{Time axis alignment}
The pulse compression process by using the time-reversed version of
an FVN does not correctly work
when the sampling clock of the DA conversion and the AD conversion does not
share the same master clock.
The test signals made from FVN sequence provide a way to
align the original time axis (for DA conversion) and the
time axis of the recorded signal (by AD conversion).
It is because the test signals based on FVN sequences are periodic.

Let $f_\mathrm{DA}(t)$ and $f_\mathrm{AD}(t)$ represent
the instantaneous (fundamental) frequencies of the DA-signal and the AD-signal, respectively.
Following equation provides the phase of the fundamental component of each signal.
Let $\varphi_\mathrm{DA}(t)$ and $\varphi_\mathrm{AD}(t)$ represent
the fundamental phase of each signal.
\begin{align}
\varphi_\mathrm{DA}(t) & = 2\pi\int_{0}^{t} f_\mathrm{DA}(\tau) d\tau \\
\varphi_\mathrm{AD}(t) & = 2\pi\int_{0}^{t} f_\mathrm{AD}(\tau) d\tau ,
\end{align}

Let $t_\mathrm{DA}(t_\mathrm{AD})$ represents the time alignment function which converts the AD-time axis to the DA-time axis. 
The following equation defines it:
\begin{align}
t_\mathrm{DA}(t_\mathrm{AD}) & = \varphi_\mathrm{DA}^{-1}(\varphi_\mathrm{AD}(t_\mathrm{AD})) ,
\end{align}
where $\varphi_\mathrm{DA}^{-1}(\varphi)$ represents the inverse function of $\varphi_\mathrm{DA}(t)$.

A simple procedure based on the analytic signal $h(t; f_\mathrm{o})$ made from the six-term cosine series and
the complex exponential function provides accurate estimates of instantaneous frequency.
\begin{align}
h(t; f_\mathrm{o}) & = \exp(2\pi j f_\mathrm{o} t) \sum_{k=0}^5 a_k \cos\left(\frac{2\pi c_\mathrm{mag} k f_\mathrm{o} t}{6}\right) ,
\end{align}
where the coefficients $\left\{a_k\right\}_{k=0}^5$ is the same as the phase manipulation function (4).
The coefficient $c_\mathrm{mag}$ is for slightly stretching the envelope of the analytic signal $h(t; f_\mathrm{o})$.
Using this analytic signal for the impulse response of a complex band-pass filter, it selects the
fundamental component of the test signal and outputs an analytic signal $y(t)$.
For the discrete version of the time signal $y[n]$ with the sampling frequency $f_s$,
the following equation provides the instantaneous frequency $f_i[n]$.
\begin{align}
f_i[n] & = \angle\left[\frac{y[n+1]}{y[n]} \right]\frac{f_s}{2\pi} ,
\end{align}
where $\angle[c]$ represents the angle of the complex number $c$
and $n$ represents the index of the discrete-time.
This procedure provides $\varphi_\mathrm{DA}(t)$ and $\varphi_\mathrm{AD}(t)$ in the measurement.

\begin{figure}[tbp]
\begin{center}
\includegraphics[width=0.48\hsize]{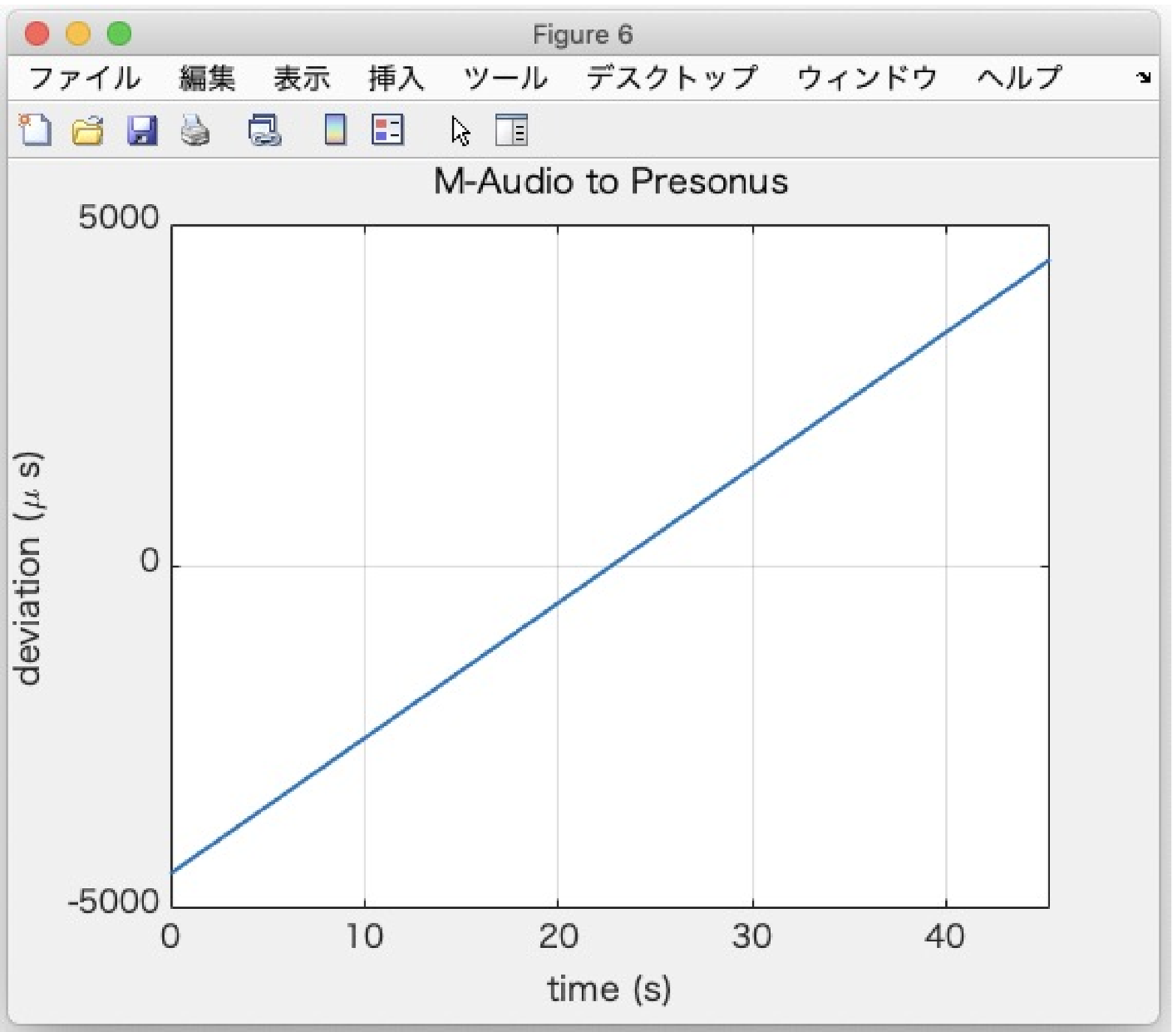} 
\hfill
\includegraphics[width=0.48\hsize]{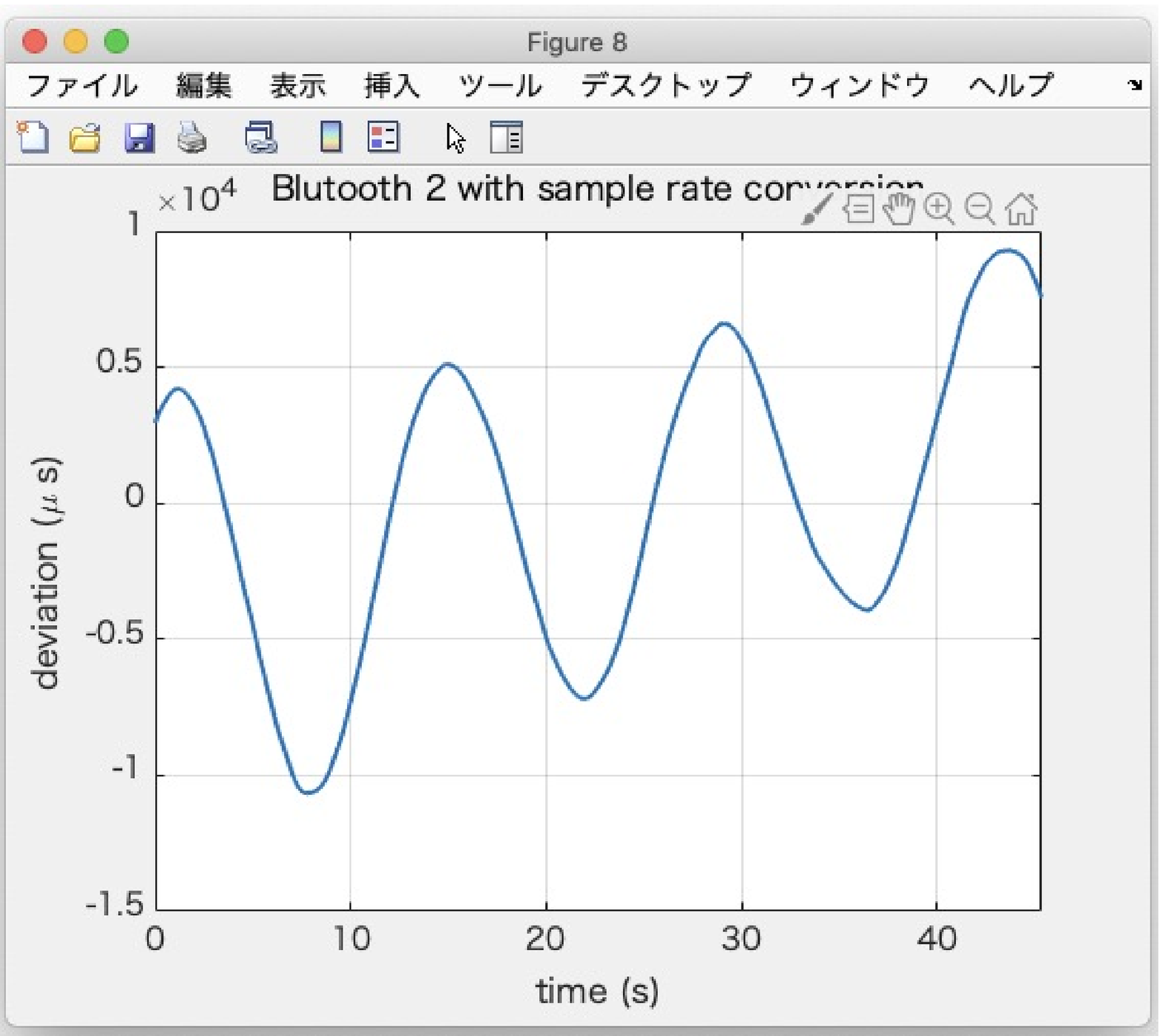} \\
\vspace{3mm}
\includegraphics[width=0.48\hsize]{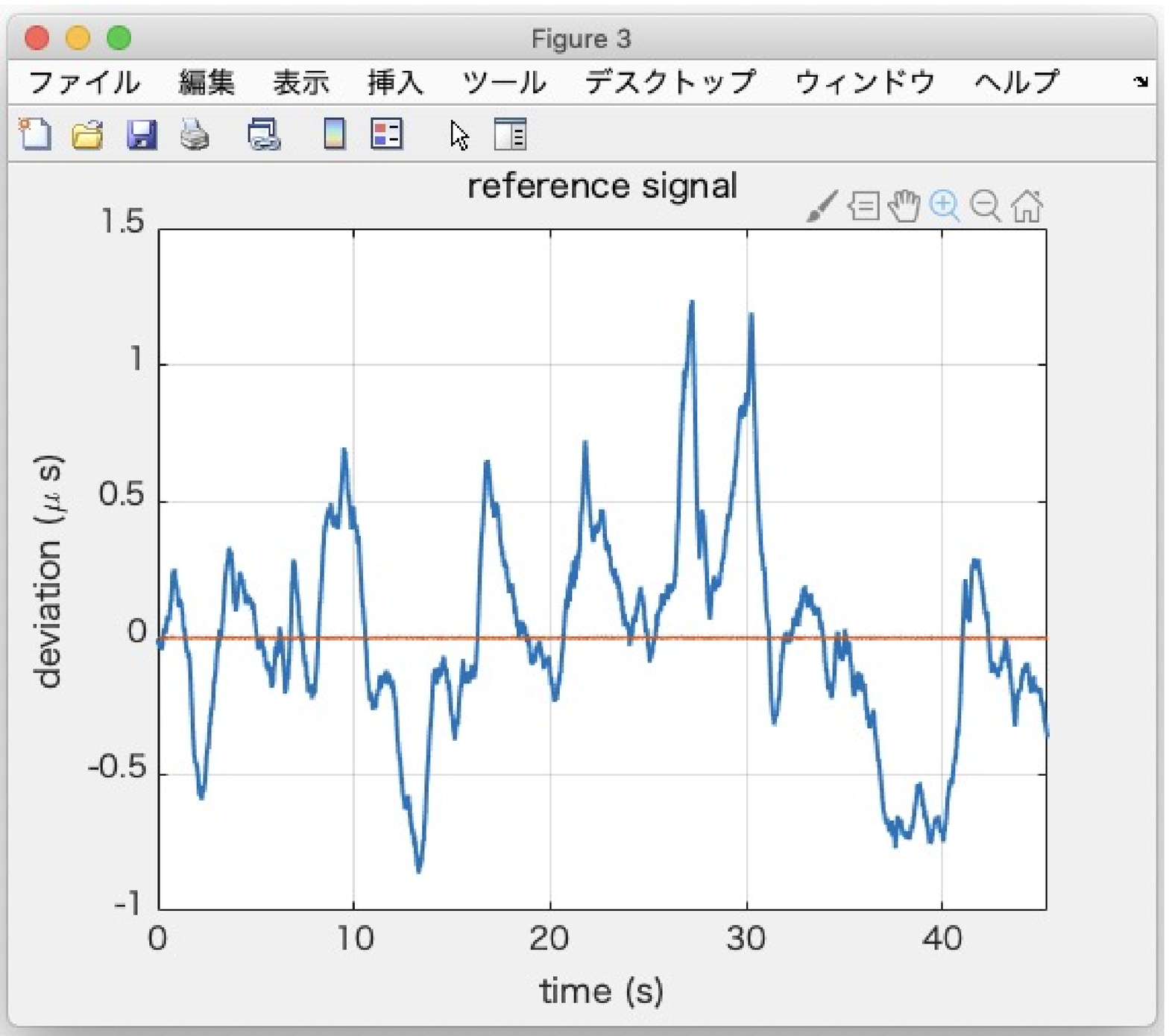} 
\hfill
\includegraphics[width=0.48\hsize]{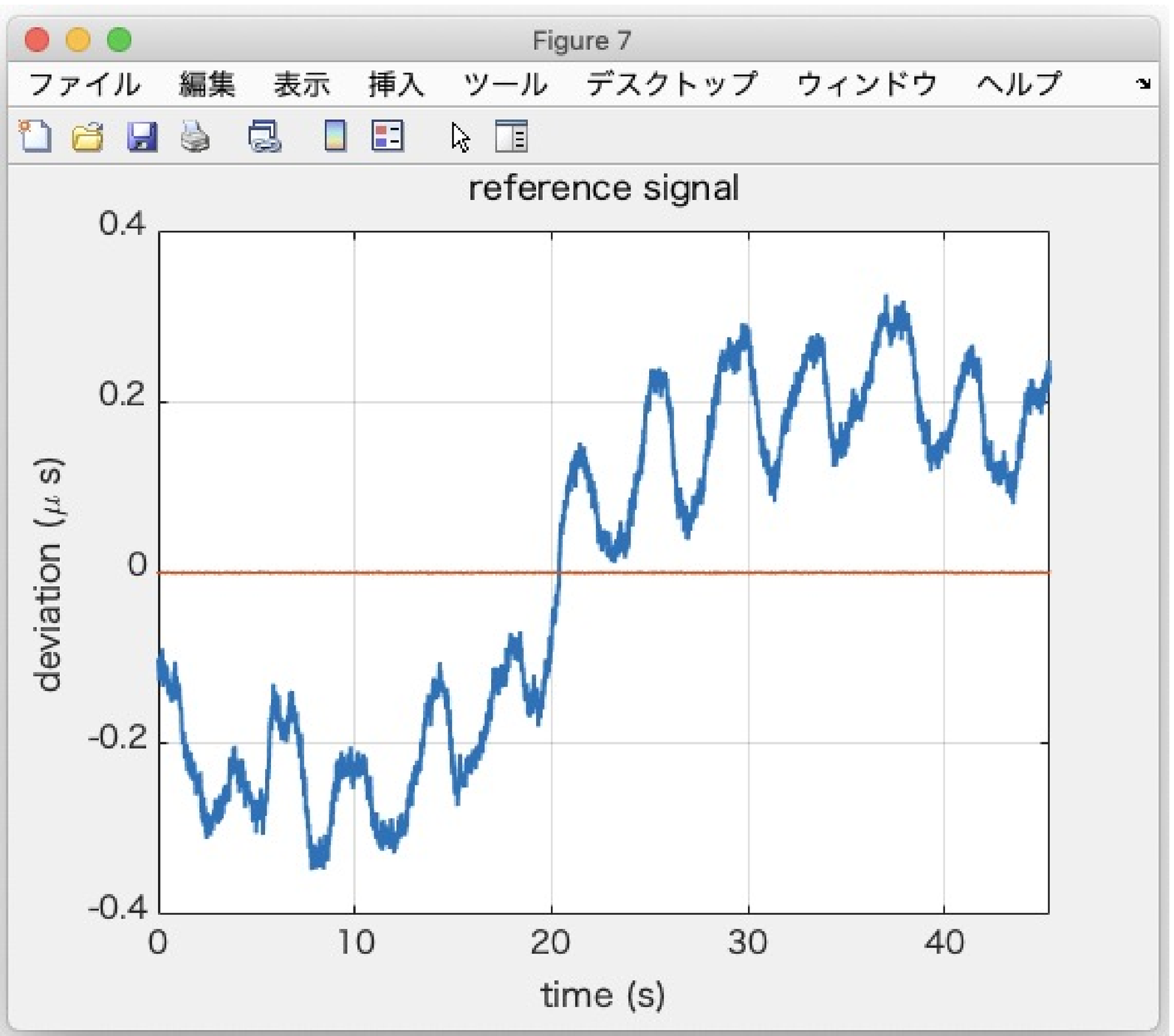} \\
\caption{Time axis deviations from the identity mapping.\label{fig:timedev}}
\end{center}
\end{figure}

Figure~\ref{fig:timedev} shows examples of time alignment.
The alignments are very close to the identity mapping.
The plots use deviations from the identity mapping to show them clearly.
The top left plot shows the deviation of two
different audio interfaces (PreSonus STUDIO $2|6$ USB and M-AUDIO M-TRACK 2X2M).
The top right plot shows the deviation using a Bluetooth connection to a powered loudspeaker.
The bottom two plots show the results using the same sampling clock for DA and AD conversion.
These plots show deviations caused by the modification of the propagation delay from the
loudspeaker to the microphone (50~cm apart).
In the left plot, blowing the path between the loudspeaker and the microphone modulated the propagation delay.
In the right plot, breathing modulated the propagation delay even though the experimenter was 1~m away from the microphone and the
loudspeaker.

\end{document}